\documentclass[12pt]{article}
\usepackage{amsmath,amsfonts,amssymb,amsthm,amsbsy,euscript}

\newtheorem*{claim}{Claim}
\theoremstyle{remark}
\newtheorem*{rem}{Remark}
\newtheorem{example}{Example}

\newcommand{\schouten}[1]{[\![#1]\!]}
\newcommand{\Or}{{\rm O\oldvec{r}}}
\newcommand{\cP}{\mathcal{P}}
\newcommand{\cQ}{\mathcal{Q}}

\newcommand{\cX}{{\EuScript X}}    
\newcommand{\Edges}{\mathsf{E}}
\newcommand{\dd}{\partial}
\DeclareMathOperator{\Id}{d}
\DeclareMathOperator{\im}{im}

\newcommand{\lshad}{[\![}
\newcommand{\rshad}{]\!]}

\def\oldvec{\mathaccent "017E\relax }

\newcommand{\by}[1]{\textrm{{#1}}}
\newcommand{\jour}[1]{\textrm{{#1}}}
\newcommand{\vol}[1]{\textrm{{V.~#1.}}}
\newcommand{\book}[1]{\textrm{{#1}}}

\newcommand{\coloneqq}{\mathrel{{:}{=}}}

\title{Poisson brackets symmetry from the\\ pentagon\/-\/wheel cocycle in the graph complex}

\author{R. Buring\thanks{%
Mathematical Institute, 
Johannes Gutenberg University of Mainz, 
Staudingerweg~9, 
\mbox{D-\/55128} 
Germany.},%
$^{,\ddagger}$
\quad 
A.\,V.\,Kiselev\thanks{%
Johann Bernoulli Institute for Mathematics \& 
Computer Science, University of Groningen, P.O.~Box~407, 9700~AK Groningen, The~Netherlands.
Partially supported 
by JBI~RUG project~103511 (Groningen). 
A~part of this research was done while R.\,B.\ and A.V.K.\ 
were visiting at the IH\'ES (Bures\/-\/sur\/-\/Yvette, France) 
and A.V.K.\ was visiting at the University of Mainz.
},$^{,\S}$
\quad
N.\,J.\,Rutten$^{\dagger}$}

\date{{E-mail}:\quad 
${}^{\ddagger}$\:\texttt{rburing\symbol{"40}uni-mainz.de},\quad
${}^{\S}$\:
\texttt{A.V.Kiselev\symbol{"40}rug.nl}%
}

\begin{document}
\maketitle

\begin{abstract}\noindent%
Kontsevich designed a scheme to generate infinitesimal symmetries $\dot{\cP} = \cQ(\cP)$ of Poisson brackets $\cP$ on all affine manifolds $M^r$; every such deformation is encoded by 
ori\-en\-ted graphs on $n+2$ vertices and $2n$ edges.
In particular, 
these symmetries can be ob\-tai\-ned by orienting 
sums of non\/-\/oriented graphs $\gamma$ on $n$ vertices and $2n-2$ edges.
The bi\/-\/vector flow $\smash{\dot{\cP}} = \Or(\gamma)(\cP)$ preserves the space of Poisson structures if 
$\gamma$ is a cocycle with respect to the vertex\/-\/expanding differential in the 
graph complex.

A class of such cocycles $\boldsymbol{\gamma}_{2\ell+1}$
is known to exist: marked by $\ell \in \mathbb{N}$, each of them contains a $(2\ell+1)$-gon wheel with a nonzero coefficient.
At $\ell=1$ the tetrahedron $\boldsymbol{\gamma}_3$
itself is a cocycle; at $\ell=2$ the Kontsevich\/--\/Willwacher pentagon\/-\/wheel cocycle $\boldsymbol{\gamma}_5$ consists of two graphs.
We reconstruct the symmetry 
$\cQ_5(\cP) = \Or(\boldsymbol{\gamma}_5)(\cP)$ and verify that $\cQ_5$~is a Poisson cocycle indeed: $\lshad\cP,\cQ_5(\cP)\rshad\doteq 0$ via $\lshad\cP,\cP\rshad=0$.
\end{abstract}

\noindent%
Generic classical Poisson brackets $\cP$ can be deformed along no less than countably many directions
(in the spaces of bi\/-\/vectors) such that they stay Poisson at least infinitesimally and the change of brackets is not necessarily induced by a diffeomorphim along integral curves 
of a 
vector field on the Poisson manifold at hand.%
\footnote{The dilation $\smash{\dot{\cP}} = \cP$ is an 
example of symmetry for 
Jacobi identity
; we 
study nonlinear flows $\smash{\dot{\cP}} = \cQ(\cP)$ which are
universal w.r.t.\ 
all affine manifolds 
and should 
persist under the quantization $\frac{\hbar}{\boldsymbol i} \{\cdot,\cdot\}_\cP \mapsto [\cdot, \cdot]$.
}
The 
use of graphs converts this infinite analytic problem into a set of finite combinatorial problems of
finding cocycles $\gamma \in \ker \Id$ in the 
graph complex and orienting them: $\cQ(\cP) = \Or(\gamma)(\cP)$, see the diagram.
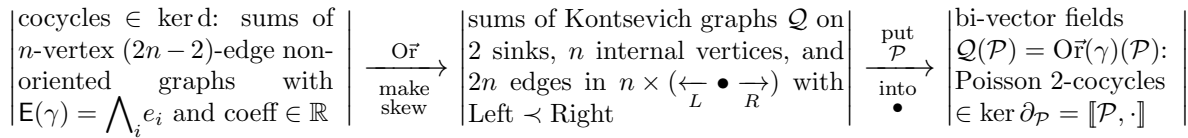
\[
\text{\raisebox{0.5mm}[8mm][5mm]{$%
\left| 
\text{\parbox{42mm}{\footnotesize%
cocycles $\in\ker\Id$: sums of $n$-vertex $(2n-2)$-edge~non- oriented graphs with $\Edges(\gamma) = \smash{\bigwedge\nolimits_{i}} 
e_i$ and coeff${}\in\mathbb{R}$ 
}
} \right|%
\xrightarrow[ \substack{ 
\text{make} \\ \text{skew} 
}]{\ \ \Or\ \ }%
\left| 
\text{\parbox{50mm}{\footnotesize%
sums of Kontsevich graphs $\cQ$ on $2$ sinks, $n$ internal vertices, and $2n$ edges in 
$\smash{n \times (\xleftarrow[L]{} {\bullet} \xrightarrow[R]{})}$ with Left${}\prec{}$Right
}
}
\right|%
\xrightarrow[ \substack{ 
\text{into} \\ \bullet
}]{\ \substack{ 
\text{put} \\ \cP 
}\ }
\left| 
\text{\parbox{28.5mm}{\footnotesize%
bi-vector fields\\ $\cQ(\cP)=\Or(\gamma)(\cP)$:\\ Poisson $2$-cocycles\\ $\in \ker \partial_\cP = \schouten{\cP, \cdot}$}
}
\right| 
$}}
\]

\noindent\textbf{1. Graph complex theory.}
There are several ways to introduce a differential on the space of non\/-\/oriented graphs 
(see \cite{
Ascona96,
KhoroshkinWillwacherZivkovic}
).
We consider the real vector space of finite non\/-\/oriented graphs such that each of them is equipped with a wedge product of edges, i.e.\ we suppose that for every graph its edges $e_i$ are enumerated $I$,\ $II$,\ $\ldots$ and proclaimed parity\/-\/odd, so that $\Edges(\gamma) \coloneqq \bigwedge_i e_i$ and $(\gamma$,\ $I \wedge II \wedge III \wedge \ldots) = -(\gamma$,\ $ II \wedge I \wedge III \wedge \ldots)$,~etc.

Suppose also that all 
vertices are at least tri\/-\/valent (cf.~\cite{JNMP17,WillwacherGRT}).
On this subspace (which we study here), 
the differential 
amounts to a blow\/-\/up --\,via the Leibniz rule\,-- of vertices in a graph $\gamma$; every vertex $v$ at hand is replaced by the new edge $E$ such that every edge which was incident to $v$ in $\gamma$ is now re-\/directed to one of the two ends of $E$.
The choice where to direct a given edge does not depend on a similar choice for other such edges, but overall, the valency of either end of $E$ must be at least two.
\footnote{In earnest, 
graphs with valency $1$ 
of an end of $E$ 
cancel out 
in the action 
of this differential $\Id$, cf.~\cite{JNMP17,KhoroshkinWillwacherZivkovic}.}
By construction, the new edge $E$ is placed firstmost
in the wedge product of edges in every graph $g$ in $\Id(\gamma)$: whenever $\Edges(\gamma) = I \wedge II \wedge \ldots$, let $\Edges(g) = E \wedge I \wedge II \wedge \ldots$.
Now one has~$\Id^2 = 0$.

\begin{example}
\label{ExDifferential}
Let $w_4\mathrel{{:}{=}}$\raisebox{-4pt}[0pt][0pt]{
{\unitlength=0.8mm
\begin{picture}(7,7)(-3.5,-3.5)
\put(3.5,3.5){\circle*{1}}
\put(-3.5,3.5){\circle*{1}}
\put(-3.5,-3.5){\circle*{1}}
\put(3.5,-3.5){\circle*{1}}
\put(0,0){\circle*{1.2}}
\put(3.5,3.5){\line(-1,0){7}}
\put(-3.5,3.5){\line(0,-1){7}}
\put(-3.5,-3.5){\line(1,0){7}}
\put(3.5,-3.5){\line(0,1){7}}
\put(3.5,3.5){\line(-1,-1){7}}
\put(3.5,-3.5){\line(-1,1){7}}
\end{picture}
}}
, and let the edge ordering in these graphs 
be lexicographic:
\[
\delta_6 \coloneqq \text{\raisebox{0pt}[5mm][3mm]{
$\Id\Biggl( \phantom{mll}$
\raisebox{3pt}[1pt][1pt]{
{\unitlength=0.09mm
\begin{picture}(0,0)(0,60)
\put(0,0){\circle*{10}}
\put(0,0){\line(0,1){120}}
\put(0,120){\circle*{10}}
\put(0,120){\line(1,0){120}}
\put(120,120){\circle*{10}}
\put(0,0){\line(1,0){120}}
\put(120,0){\circle*{10}}
\put(120,0){\line(0,1){120}}
\put(60,60){\circle*{10}}
\qbezier[60](60,60)(90,90)(120,120)
\qbezier[60](60,60)(30,90)(0,120)
\qbezier[60](60,60)(90,30)(120,0)
\qbezier[60](60,60)(30,30)(0,0)
\put(-60,60){\circle*{10}}
\qbezier[60](-60,60)(-30,90)(0,120)
\qbezier[60](-60,60)(-30,30)(0,0)
\put(-60,60){\line(1,0){120}}
\put(-25,-10){\tiny$5$}
\put(-25,115){\tiny$1$}
\put(131,115){\tiny$4$}
\put(131,-10){\tiny$2$}
\put(73,53){\tiny$6$}
\put(-85,53){\tiny$3$}
\end{picture}
}
}
$\phantom{mmml}
\Biggr)$
}}
= 
2
\phantom{mm}
\raisebox{3pt}[1pt][1pt]{
{\unitlength=0.09mm
\begin{picture}(0,0)(0,60)
\put(0,0){\circle*{10}}
\put(0,0){\line(0,1){120}}
\put(0,120){\circle*{10}}
\put(0,120){\line(1,0){120}}
\put(120,120){\circle*{10}}
\put(0,0){\line(1,0){120}}
\put(120,0){\circle*{10}}
\put(120,0){\line(0,1){120}}
\put(30,60){\circle*{10}}
\put(30,60){\line(1,0){60}}
\put(90,60){\circle*{10}}
\qbezier(0,120)(15,90)(30,60)
\qbezier(0,0)(15,30)(30,60)
\qbezier(120,0)(105,30)(90,60)
\qbezier(120,120)(105,90)(90,60)
\put(-60,60){\circle*{10}}
\qbezier[60](-60,60)(-30,90)(0,120)
\qbezier[60](-60,60)(-30,30)(0,0)
\put(-60,60){\line(1,0){90}}
\put(-25,-10){\tiny$7$}
\put(-25,115){\tiny$5$}
\put(130,115){\tiny$3$}
\put(130,-10){\tiny$1$}
\put(39,66){\tiny$6$}
\put(103,53){\tiny$2$}
\put(-85,53){\tiny$4$}
\end{picture}
}
}
\phantom{mmll}
{} + 4
\phantom{mm}
\raisebox{3pt}[1pt][1pt]{
{\unitlength=0.09mm
\begin{picture}(0,0)(0,60)
\put(0,0){\circle*{10}}
\put(0,0){\line(0,1){120}}
\put(0,120){\circle*{10}}
\put(0,120){\line(1,0){120}}
\put(120,120){\circle*{10}}
\put(0,0){\line(1,0){120}}
\put(120,0){\circle*{10}}
\put(120,0){\line(0,1){120}}
\put(60,60){\circle*{10}}
\qbezier[60](60,60)(90,90)(120,120)
\qbezier[60](60,60)(30,90)(0,120)
\qbezier[60](60,60)(90,30)(120,0)
\qbezier[60](60,60)(30,30)(0,0)
\put(-60,60){\circle*{10}}
\qbezier[60](-60,60)(-30,90)(0,120)
\qbezier[60](-60,60)(-30,30)(0,0)
\put(30,90){\circle*{10}}
\put(37,90){\tiny$2$}
\qbezier[60](-60,60)(-15,75)(30,90)
\put(-25,-10){\tiny$7$}
\put(-25,115){\tiny$5$}
\put(131,115){\tiny$1$}
\put(131,-10){\tiny$4$}
\put(73,53){\tiny$6$}
\put(-85,53){\tiny$3$}
\end{picture}
}
}
\phantom{mmm}
+ 4
\phantom{mm}
\raisebox{3pt}[1pt][1pt]{
{\unitlength=0.09mm
\begin{picture}(0,0)(0,60)
\put(0,0){\circle*{10}}
\put(0,0){\line(0,1){120}}
\put(0,120){\circle*{10}}
\put(0,120){\line(1,0){120}}
\put(120,120){\circle*{10}}
\put(0,0){\line(1,0){120}}
\put(120,0){\circle*{10}}
\put(120,0){\line(0,1){120}}
\put(60,60){\circle*{10}}
\qbezier[60](60,60)(90,90)(120,120)
\qbezier[60](60,60)(30,90)(0,120)
\put(-60,60){\circle*{10}}
\qbezier[60](-60,60)(-30,90)(0,120)
\qbezier[60](-60,60)(-30,30)(0,0)
\put(-60,60){\line(1,0){120}}
\qbezier[60](0,0)(30,15)(60,30)
\qbezier[60](120,0)(90,15)(60,30)
\put(60,30){\circle*{10}}
\put(60,30){\line(0,1){30}}
\put(-25,-10){\tiny$6$}
\put(-25,115){\tiny$7$}
\put(131,115){\tiny$3$}
\put(131,-10){\tiny$1$}
\put(73,53){\tiny$5$}
\put(-85,53){\tiny$4$}
\put(30,30){\tiny$2$}
\end{picture}
}
}
\phantom{mmll}
- 4
\phantom{mm}
\raisebox{3pt}[1pt][1pt]{
{\unitlength=0.09mm
\begin{picture}(0,0)(0,60)
\put(0,0){\circle*{10}}
\qbezier[60](0,0)(-15,45)(-30,90)
\put(0,120){\circle*{10}}
\put(0,120){\line(1,0){120}}
\put(120,120){\circle*{10}}
\put(0,0){\line(1,0){120}}
\put(120,0){\circle*{10}}
\put(120,0){\line(0,1){120}}
\put(60,60){\circle*{10}}
\qbezier[60](60,60)(90,90)(120,120)
\qbezier[60](60,60)(30,90)(0,120)
\qbezier[60](60,60)(90,30)(120,0)
\qbezier[60](60,60)(30,30)(0,0)
\put(-60,60){\circle*{10}}
\qbezier[60](-60,60)(-30,90)(0,120)
\qbezier[60](-60,60)(-30,30)(0,0)
\put(-60,60){\line(1,0){120}}
\put(-30,90){\circle*{10}}
\put(-25,-10){\tiny$6$}
\put(-25,115){\tiny$2$}
\put(131,115){\tiny$1$}
\put(131,-10){\tiny$5$}
\put(73,53){\tiny$7$}
\put(-85,53){\tiny$4$}
\put(-55,85){\tiny$3$}
\end{picture}
}
}\phantom{mmm}
\]
A flip over a diagonal in $w_4$ swaps three pairs of edges; $3$ is odd, so by this symmetry, $\Edges(w_4) = -\Edges(w_4)$, i.e.\ 
$w_4$ is a \emph{zero} graph.\footnote{One proves that $\Id$(zero graph) $=$ sum of zero graphs and graphs with zero coefficients.} 
By this, $\Id(w_4) = 0$.
Because $\Id^2 = 0$, one has $\Id(\delta_6) = 0$ for the coboundary $\delta_6 \in \im \Id$.
Put $\boldsymbol{\gamma}_3\mathrel{{:}{=}}$
\raisebox{-4.75pt}[0pt][0pt]{
{\unitlength=0.75mm
\begin{picture}(7,7)(-3.5,-3.5)
\put(3.5,3.5){\circle*{1}}
\put(-3.5,3.5){\circle*{1}}
\put(-3.5,-3.5){\circle*{1}}
\put(3.5,-3.5){\circle*{1}}
\put(3.5,3.5){\line(-1,0){7}}
\put(-3.5,3.5){\line(0,-1){7}}
\put(-3.5,-3.5){\line(1,0){7}}
\put(3.5,-3.5){\line(0,1){7}}
\qbezier[30](3.5,3.5)(2,2)(0.75,0.75)
\qbezier[30](-3.5,-3.5)(-2,-2)(-1,-1)
\put(3.5,-3.5){\line(-1,1){7}}
\end{picture}
}}%
; another example of \emph{non}trivial cocycle, $\boldsymbol{\gamma}_5\not\in \im \Id$, also on $n$ vertices and $2n-2$ edges, is given on p.~\pageref{FigPentagon}.
\end{example}

The notion of \emph{oriented} Kontsevich graphs from~\cite{Ascona96} 
was recalled in~\cite{f16,sqs15,JPCS17}.
Every such graph is built over $m$ ordered sinks from $n$ wedges $\smash{\xleftarrow{L}{\bullet}\xrightarrow{R}}$: each top $\bullet$ of the wedge is the source of exactly two arrows (which 
are ordered by Left${}\prec{}$Right).
Let $(M^r$,\ $\cP)$ be a real affine Poisson manifold of dimension $r$; let $x^1$,\ $\ldots$,\  $x^r$ be local coordinates.
By decorating each edge with its own summation index that runs from $1$ to $r$, by identifying every such edge $\smash{\xrightarrow{i}}$ with $\partial/\partial x^i$ acting on the content of arrowhead vertex, and by placing a copy of the Poisson bi-\/vector $\cP = (\cP^{ij}$) at the top $\bullet$ of each wedge $\smash{\xleftarrow{i} {\bullet} \xrightarrow{j}}$, we associate a polydifferential operator (e.g., an $m$-\/vector) with every such graph.
The arguments of the operator are contained in the $m$ respective sinks. 
The resulting polydifferential operators are differential\/-\/polynomial in the coefficients $\cP^{ij}$ of a given Poisson structure $\cP$.
It is known that for $\cP$ Poisson (hence $\schouten{\cP,\cP} = 0$ under the Schouten bracket), its adjoint action 
$\partial_\cP \coloneqq \schouten{\cP, \cdot}$ is a differential on the space of multi\/-\/vectors.
One can try finding Poisson cohomology cocycles $\cQ\in\ker \partial_\cP$ by assuming they are realized using the Kontsevich oriented graphs.

Now let us note that certain sums $\cQ$ of oriented graphs built on two sinks from $n$ wedges can be obtained by taking all 
admissible ways to orient graphs $\gamma$ on $n$ vertices and $2n-2$ edges (clearly, 
two sinks and two edges 
into them are added).
Moreover, suppose that $\gamma \in \ker \Id$ in vertex\/-\/edge bi-\/grading $(n,2n-2)$ is such that this sum of graphs can be oriented to 
yield a sum of Kontsevich graphs on two sinks, $n$ internal vertices 
and $2n$ edges.
Then, in fact, these oriented graphs, taken with suitable coefficients $\in\mathbb{R}$, do assemble to a Poisson cocycle $\cQ(\cP) \in \ker \partial_\cP$.
Let this orientation mapping be denoted by $\Or$ 
(cf.~\cite{Ascona96} and~\cite{f16,JPCS17}).
\footnote{The present paper is aimed to help us reveal the general formula of the morphism $\Or$ which connects the two graph complexes.} 


\noindent\textbf{2. The pentagon\/-\/wheel cocycle.}\label{SecPentagon}
The mechanism of factorization $\schouten{\cP, \cQ(\cP)} \doteq 0$ via $\schouten{\cP,\cP} = 0$ for the cocycle condition $\cQ(\cP) \in \ker \partial_\cP$ is known from~\cite{sqs15}, where it is used in a similar problem of the $\star$-product associativity (cf.~\cite{cpp}).
In~\cite{f16} this mechanism is applied to the Kontsevich tetrahedral flow $\cQ_3(\cP) = \Or(\boldsymbol{\gamma}_3)(\cP)$.
%
Would the mapping $\Or$ be known, the verification $\Or(\gamma) \in \ker \partial_\cP$ 
is still compulsory (e.g., by using a factorization via the Jacobi identity for $\cP$).
But for us now, the factorization $\schouten{\cP, \cQ_5(\cP)}= \Diamond\bigl(\cP, \schouten{\cP,\cP}\bigr)$ is the way to find the right formula of the flow $\dot{\cP} = \cQ_5(\cP)$ 
that should correspond to the Kontsevich\/--\/Willwacher pentagon-wheel cocycle $\boldsymbol{\gamma}_5$ under the orientation mapping,
$\cQ_5 = \Or(\boldsymbol{\gamma}_5)$,
giving one 
solution $\cQ_5$ yet not necessarily unique operator~$\Diamond$. 

\begin{example}
\label{Ex3wheel}
There are only two essentially different admissible ways to orient (and skew\/-\/symmetrize with respect to sinks) the tetrahedron 
$\boldsymbol{\gamma}_3 \in \ker \Id$. 
Each of the three oriented graphs in the flow $\cQ_3$ is encoded by the list of targets for the ordered pair of edges issued from the $i$th vertex ($m = 2 \leqslant i \leqslant 5 = m+n-1$), and a 
coefficient $\in\mathbb{Z}$.
Specifically, we have that $\cQ_3=
1 \cdot( \mathsf{0},\mathsf{1} ; \mathsf{2},\mathsf{4} ; \mathsf{2},\mathsf{5} ; \mathsf{2},\mathsf{3})
-3 \cdot( \mathsf{0},\mathsf{3} ; \mathsf{1},\mathsf{4} ; \mathsf{2},\mathsf{5} ; \mathsf{2},\mathsf{3}  + 
   \mathsf{0},\mathsf{3} ; \mathsf{4},\mathsf{5} ; \mathsf{1},\mathsf{2} ; \mathsf{2},\mathsf{ 4} )   
$; 
the analytic formula of the respective bi-\/differential operators acting on the sinks 
content~$f$, $g$ is 
$\cQ_3(f,g)=
\dd_{kmp}\cP^{ij} \dd_q\cP^{k\ell} \dd_\ell\cP^{mn} \dd_n\cP^{pq}\cdot \dd_if \dd_j g
-3 \dd_{mp}\cP^{ij} \dd_{jq}\cP^{k\ell} \dd_\ell\cP^{mn} \dd_n\cP^{pq}\cdot \dd_if \dd_k g
-3 \dd_{np}\cP^{ij} \dd_j\cP^{k\ell} \dd_{kq}\cP^{mn} \dd_\ell\cP^{pq}\cdot \dd_i f \dd_m g
$.
A factorization of $\schouten{\cP, \cQ_3(\cP)}$ via $8$ tri\/-\/vector graphs containing $\schouten{\cP,\cP}$ is explained in~\cite{f16}, based on~\cite{sqs15}.
\end{example}

\hangindent=-6.5cm\hangafter=-3%
{\unitlength=1mm
\begin{picture}(0,0)(-32,2.5)
\put(65,0){$\boldsymbol{\gamma}_5 = {}$}
\put(85,0){
{\unitlength=0.3mm
\begin{picture}(55,53)(5,-5)
\put(27.5,8.5){\circle*{3}}
\put(0,29.5){\circle*{3}}
\put(-27.5,8.5){\circle*{3}}
\put(-17.5,-23.75){\circle*{3}}
\put(17.5,-23.75){\circle*{3}}
\qbezier
(27.5,8.5)(0,29.5)(0,29.5)
\qbezier
(0,29.5)(-27.5,8.5)(-27.5,8.5)
\qbezier
(-27.5,8.5)(-17.5,-23.75)(-17.5,-23.75)
\qbezier
(-17.5,-23.75)(17.5,-23.75)(17.5,-23.75)
\qbezier
(17.5,-23.75)(27.5,8.5)(27.5,8.5)
\put(0,0){\circle*{3}}
\qbezier
(27.5,8.5)(0,0)(0,0)
\qbezier
(0,29.5)(0,0)(0,0)
\qbezier
(-27.5,8.5)(0,0)(0,0)
\qbezier
(-17.5,-23.75)(0,0)(0,0)
\qbezier
(17.5,-23.75)(0,0)(0,0)
\end{picture}
}
}
\put(95,0){${}+\dfrac{5}{2}$}
\put(114,0){
{\unitlength=0.4mm
\begin{picture}(50,30)(0,-4)
\put(12,0){\circle*{2.5}}
\put(-12,0){\circle*{2.5}}
\put(25,15){\circle*{2.5}}
\put(-25,15){\circle*{2.5}}
\put(-25,-15){\circle*{2.5}}
\put(25,-15){\circle*{2.5}}
\put(-12,0){\line(1,0){24}}
\put(-25,15){\line(1,0){50}}
\put(-25,-15){\line(1,0){50}}
\put(-25,-15){\line(0,1){32}}
\put(25,15){\line(0,-1){32}}
\qbezier
(25,15)(12,0)(12,0)
\qbezier
(-25,15)(-12,0)(-12,0)
\qbezier
(-25,-15)(-12,0)(-12,0)
\qbezier
(25,-15)(12,0)(12,0)
\put(-12.5,17){\oval(25,10)[t]}
\put(12.5,-17){\oval(25,10)[b]}
\put(0,2){\line(0,1){11}}
\put(0,-2){\line(0,-1){11}}
\end{picture}
}%
}
\end{picture}%
}%
\label{FigPentagon}%
Now consider the pentagon\/-\/wheel cocycle
$\boldsymbol{\gamma}_5 \in \ker \Id$,
see~\cite{JNMP17}.
By orienting both graphs in $\boldsymbol{\gamma}_5$ (i.e.\ by shifting the vertex labelling by $+1 = m-1$, adding two edges to the sinks $\mathsf{0}$,\ $\mathsf{1}$, and keeping only those oriented graphs out of $1024 = 2^{\text{\#edges}}$ which are built from $\xleftarrow{}{\bullet} \xrightarrow{}$) and skew\/-\/symmetrizing with respect to $\mathsf{0} \rightleftarrows \mathsf{1}$, we obtain $91$ parameters for 
Kontsevich graphs on $2$ sinks, $6$ internal vertices, and $12$ ($=6$ pairs) of edges.
We take the sum $\cQ$ of these $91$ bi-\/vector graphs (or skew differences of 
Kontsevich graphs)
with their undetermined coefficients, and for the set of tri\/-\/vector graphs occurring in $\schouten{\cP,\cQ}$, we generate all the possibly 
needed tri\/-\/vector ``Leibniz'' graphs with $\schouten{\cP,\cP}$ inside.\footnote{%
The algorithm from~\cite[\S1.2]{JPCS17} produces 41031 
Leibniz graphs in $\nu=3$ 
iterations and 56509 
at~$\nu\geqslant7$.}
This yields 41031 
such Leibniz graphs,
which, with undetermined coefficients, provide the ansatz for the r.-h.s.\ of the factorization problem
\label{EqFactor}
$\schouten{\cP,\cQ(\cP)} = \Diamond\bigl(\cP,\schouten{\cP,\cP}\bigr)$.
This gives us an inhomogeneous system of 463,344 
linear algebraic equations for both the coefficients in $\cQ$ and~$\Diamond$.
In its l.-h.s., we fix the coefficient of one bi\/-\/vector graph\footnote{This is done because it is anticipated that, counting the number of ways to obtain a given bi\/-\/vector while orienting the nonzero cocycle~$\boldsymbol{\gamma}_5$, none of the coefficients in a solution~$\cQ_5$ vanishes.}
by setting it to~${\mathbf{+2}}$. 

\begin{claim}For~$\boldsymbol{\gamma}_5$,
the factorization problem $\schouten{\cP,\cQ(\cP)} = \Diamond(\cP,\schouten{\cP,\cP})$ has a solution $(\cQ_5, \Diamond_5)$\textup{;} the sum $\cQ_5$ of $167$ Kontsevich graphs \textup{(}on $m=2$ sinks $\mathsf{0},\mathsf{1}$ and $n=6$ internal vertices $\mathsf{2}$\textup{,} $\ldots$\textup{,} $\mathsf{7}$\textup{)} with integer coefficients is given in the table 
below.
\footnote{%
The analytic formula of degree\/-\/six nonlinear differential polynomial $\cQ_5(\cP)$ is given in App.~\ref{AppFormula}. 
The encoding of $8
691$ Leibniz tri\/-\/vector graphs containing the Jacobiator $\schouten{\cP,\cP}$ for the Poisson structure $\cP$ that occur in the r.-h.s.\ $\Diamond(\cP, \schouten{\cP,\cP})$ is available at \texttt{https://rburing.nl/Q5d5.txt}. 
The machine format to encode such graphs (with one tri\/-\/valent vertex for the Jacobiator) is explained 
in~\cite{JPCS17} (see also~\cite{f16,cpp}).}%
\end{claim}

{\tiny\centerline{
\begin{tabular}{l|r|}
0 1 2 4 2 5 3 6 4 7 2 4 &   $10$\\
0 1 2 4 2 5 2 6 4 7 3 4 &   $-10$\\
0 3 1 4 2 5 6 7 2 4 3 4 &   $10 $\\
0 3 4 5 1 2 6 7 2 3 3 4 &   $-10$\\
0 3 1 4 2 5 2 6 4 7 3 4 &   $10 $\\
0 3 4 5 1 2 4 6 3 7 2 3 &   $-10$\\
0 3 1 4 2 5 3 6 4 7 2 4 &   $-10$\\
0 3 4 5 1 2 2 6 3 7 3 4 &   $-10$\\
0 3 1 4 5 6 2 3 5 7 2 5 &   $-10$\\
0 3 4 5 2 6 4 7 1 2 4 6 &   $10 $\\
0 3 4 5 1 6 2 4 5 7 2 5 &   $10 $\\
0 3 4 5 2 6 4 6 1 7 2 4 &   $-10$\\
0 3 4 5 2 6 4 7 2 7 1 4 &   $-10$\\
0 3 4 5 1 6 2 4 3 7 2 3 &   $10 $\\
0 3 4 5 2 6 6 7 1 3 2 3 &   $-10$\\
0 3 4 5 2 6 2 7 1 3 3 6 &   $10 $\\
0 3 4 5 1 6 4 7 2 3 2 3 &   $-10$\\
0 3 4 5 1 5 2 6 2 7 4 5 &   $10 $\\
0 3 4 5 1 6 2 7 2 3 3 4 &   $10 $\\
0 3 4 5 1 5 2 6 4 7 2 5 &   $10 $\\
0 3 4 5 1 2 4 6 4 7 2 4 &   $-10$\\
0 3 1 4 2 5 2 6 2 7 2 3 &   $-10$\\
0 3 1 4 2 5 3 6 3 7 2 3 &   $-10$
\end{tabular}
\hskip 10pt
\begin{tabular}{l|r|}
0 3 4 5 1 2 2 6 2 7 2 4 &   $-10$\\
0 3 1 4 5 6 2 3 3 7 2 3 &   $-10$\\
0 3 4 5 2 6 2 7 1 2 2 6 &   $10$\\
0 1 2 4 2 5 2 6 2 7 2 3 &   $\mathbf{2}$\\
0 1 2 4 2 5 2 6 3 7 3 4 &   $-5$\\
0 1 2 4 2 5 3 6 3 7 2 4 &   $5 $\\
0 1 2 4 2 5 2 6 3 7 4 5 &   $-5 $\\
0 1 2 4 2 5 2 6 4 7 3 5 &   $-5 $\\
0 3 1 4 5 6 2 7 5 7 2 3 &   $5 $\\
0 3 4 5 5 6 6 7 2 7 1 2 &   $5 $\\
0 3 1 4 2 5 6 7 2 4 3 6 &   $5 $\\
0 3 4 5 1 2 6 7 2 7 3 4 &   $-5 $\\
0 3 1 4 2 5 2 6 3 7 4 5 &   $5 $\\
0 3 4 5 1 2 4 6 2 7 3 5 &   $-5 $\\
0 3 1 4 2 5 2 6 4 7 3 5 &   $5 $\\
0 3 4 5 1 2 4 6 3 7 2 5 &   $-5 $\\
0 3 4 5 1 2 6 7 2 3 4 6 &   $5 $\\
0 3 1 4 2 5 6 7 2 7 3 4 &   $5 $\\
0 3 4 5 1 2 2 6 4 7 3 5 &   $5 $\\
0 3 1 4 2 5 3 6 2 7 4 5 &   $-5 $\\
0 3 4 5 1 2 2 6 3 7 4 5 &   $5 $\\
0 3 1 4 2 5 3 6 4 7 2 5 &   $-5 $\\
0 3 4 5 2 6 6 7 1 2 3 4 &   $5 $ 
\end{tabular}
\hskip 10pt
\begin{tabular}{l|r|}
0 3 1 4 5 6 2 3 2 7 4 5 &   $5 $\\
0 3 4 5 2 6 4 7 1 2 3 6 &   $5 $\\
0 3 1 4 5 6 2 3 5 7 2 4 &   $-5 $\\
0 3 4 5 1 2 6 7 2 4 4 6 &   $-5 $\\
0 3 1 4 2 5 6 7 2 3 2 6 &   $-5 $\\
0 3 1 4 5 6 2 3 5 7 2 3 &   $-5 $\\
0 3 4 5 2 6 4 7 1 2 2 6 &   $5 $\\
0 3 1 4 2 5 6 7 2 3 3 4 &   $5 $\\
0 3 4 5 1 2 6 7 2 3 2 4 &   $-5 $\\
0 3 1 4 2 5 3 6 4 7 2 3 &   $-5 $\\
0 3 4 5 1 2 2 6 3 7 2 4 &   $-5 $\\
0 3 1 4 2 5 6 7 2 3 3 6 &   $-5 $\\
0 3 4 5 1 2 6 7 2 4 2 6 &   $-5 $\\
0 3 4 5 1 2 6 7 2 4 3 4 &   $-5 $\\
0 3 1 4 2 5 6 7 2 3 2 4 &   $5 $\\
0 3 4 5 1 2 4 6 3 7 2 4 &   $-5 $\\
0 3 1 4 2 5 2 6 4 7 2 3 &   $-5 $\\
0 1 2 4 2 5 6 7 2 7 3 4 &   $-5 $\\
0 1 2 4 2 5 3 6 2 7 4 5 &   $5 $\\
0 1 2 4 2 5 3 6 4 7 2 5 &   $5 $\\
0 1 2 4 2 5 3 6 2 7 3 5 &   $5 $\\
0 1 2 4 2 5 3 6 3 7 2 5 &   $5 $\\
0 3 4 5 1 2 4 6 2 7 4 5 &   $-5$ 
\end{tabular}
\hskip 10pt
\begin{tabular}{l|r|}
0 3 1 4 2 5 2 6 3 7 2 5 &   $5 $\\
0 3 4 5 1 2 4 6 4 7 2 5 &   $-5 $\\
0 3 1 4 2 5 2 6 2 7 3 5 &   $5 $\\
0 3 1 4 5 6 2 6 3 7 2 3 &   $-5 $\\
0 3 4 5 2 6 4 7 2 7 1 2 &   $-5 $\\
0 3 1 4 5 6 2 3 2 7 3 4 &   $-5 $\\
0 3 4 5 2 6 6 7 1 2 2 3 &   $-5 $\\
0 3 1 4 5 6 2 3 3 7 2 4 &   $-5 $\\
0 3 4 5 2 6 2 7 1 2 3 6 &   $5 $\\
0 3 1 4 2 5 3 6 2 7 3 5 &   $-5 $\\
0 3 4 5 1 2 2 6 4 7 2 5 &   $5 $\\
0 3 1 4 2 5 3 6 3 7 2 5 &   $-5 $\\
0 3 4 5 1 2 2 6 2 7 4 5 &   $5 $\\
0 3 4 5 5 6 6 7 1 2 2 6 &   $-5 $\\
0 3 1 4 5 6 2 6 2 7 2 3 &   $5 $\\
0 1 2 4 2 5 2 6 2 7 3 4 &   $-5 $\\
0 1 2 4 2 5 2 6 3 7 2 5 &   $-5 $\\
0 1 2 4 2 5 2 6 2 7 3 5 &   $-5 $\\
0 3 4 5 2 6 6 7 1 2 4 6 &   $5 $\\
0 3 1 4 5 6 2 3 2 7 2 5 &   $-5 $\\
0 3 4 5 1 2 4 6 4 7 2 3 &   $-5 $\\
0 3 1 4 2 5 2 6 2 7 3 4 &   $5 $\\
\multicolumn{2}{c|}{(\textit{see next page})}
\end{tabular}}

}

\twocolumn[
\begin{minipage}{\textwidth}
{\tiny\centerline{
\begin{tabular}{l|r|}
0 3 4 5 1 2 2 6 4 7 3 4 &   $-5 $\\
0 3 1 4 2 5 3 6 2 7 2 4 &   $-5 $\\
0 3 1 4 5 6 2 3 3 7 2 5 &   $-5 $\\
0 3 4 5 2 6 2 7 1 2 4 6 &   $5 $\\
0 3 1 4 5 6 2 7 3 7 2 3 &   $-5 $\\
0 3 4 5 2 6 6 7 2 7 1 2 &   $-5 $\\
0 3 1 4 2 5 3 6 3 7 2 4 &   $-5 $\\
0 3 4 5 1 2 2 6 2 7 3 4 &   $-5 $\\
0 3 1 4 2 5 2 6 3 7 3 4 &   $5 $\\
0 3 4 5 1 2 4 6 2 7 2 3 &   $-5 $\\
0 3 4 5 1 6 2 7 5 7 2 4 &   $-5 $\\
0 3 4 5 2 6 4 6 1 7 2 5 &   $-5 $\\
0 3 4 5 1 6 2 7 2 5 4 6 &   $5 $\\
0 3 4 5 1 6 4 7 2 5 2 3 &   $-5 $\\
0 3 4 5 1 6 2 6 2 7 4 5 &   $5 $\\
0 3 4 5 1 6 2 7 2 7 3 4 &   $5 $\\
0 3 4 5 2 6 6 7 1 7 2 3 &   $-5 $\\
0 3 4 5 1 5 6 7 2 3 2 4 &   $5 $\\
0 3 4 5 2 6 4 6 1 7 2 3 &   $-5$ 
\end{tabular}
\hskip 10pt
\begin{tabular}{l|r|}
0 3 4 5 1 5 6 7 2 4 2 6 &   $5 $\\
0 3 4 5 2 6 2 7 1 5 3 6 &   $5 $\\
0 3 4 5 1 6 2 6 3 7 2 4 &   $5 $\\
0 3 4 5 2 6 2 6 1 7 3 4 &   $-5 $\\
0 3 4 5 2 6 4 7 1 5 2 6 &   $-5 $\\
0 3 4 5 1 6 2 7 2 5 3 4 &   $5 $\\
0 3 4 5 1 6 4 7 2 5 2 6 &   $5 $\\
0 3 4 5 1 6 4 7 2 7 2 3 &   $-5 $\\
0 3 4 5 1 6 4 6 2 7 2 5 &   $5 $\\
0 3 4 5 1 6 2 7 3 5 2 4 &   $-5 $\\
0 3 4 5 2 5 6 7 1 4 2 6 &   $-5 $\\
0 3 4 5 2 6 4 7 2 7 1 3 &   $-5 $\\
0 3 4 5 2 5 6 7 1 3 2 6 &   $-5 $\\
0 3 4 5 2 6 6 7 1 7 2 4 &   $5 $\\
0 3 4 5 1 6 2 4 5 7 2 3 &   $5 $\\
0 3 4 5 2 6 6 7 2 7 1 4 &   $-5 $\\
0 3 4 5 1 6 2 4 3 7 2 5 &   $5 $\\
0 3 4 5 2 6 2 7 1 3 4 6 &   $5 $\\
0 3 4 5 2 6 6 7 1 3 2 4 &   $-5$ 
\end{tabular}
\hskip 10pt
\begin{tabular}{l|r|}
0 3 4 5 1 6 2 7 2 3 4 6 &   $-5 $\\
0 3 4 5 1 5 2 6 4 7 2 3 &   $5 $\\
0 3 4 5 1 5 2 6 2 7 3 4 &   $-5 $\\
0 3 4 5 1 6 4 7 2 3 2 6 &   $-5 $\\
0 3 4 5 1 6 2 4 2 7 4 5 &   $-5 $\\
0 3 4 5 1 6 2 7 2 7 2 4 &   $-5 $\\
0 3 4 5 1 6 2 4 5 7 2 4 &   $5 $\\
0 3 4 5 2 6 2 6 1 7 2 4 &   $-5 $\\
0 3 4 5 1 5 2 6 4 7 2 4 &   $5 $\\
0 3 4 5 1 6 2 7 2 3 2 4 &   $5 $\\
0 3 4 5 1 6 2 4 2 7 3 4 &   $5 $\\
0 3 4 5 1 6 2 6 2 7 2 4 &   $-5 $\\
0 3 4 5 1 6 2 4 3 7 2 4 &   $5 $\\
0 3 4 5 2 6 2 7 1 5 2 6 &   $5 $\\
0 3 4 5 2 6 6 7 1 3 2 6 &   $-5 $\\
0 3 4 5 2 6 2 7 1 3 2 6 &   $5 $\\
0 3 4 5 1 6 4 7 2 3 2 4 &   $-5 $\\
0 3 4 5 1 5 2 6 2 7 2 4 &   $-5 $\\
0 3 4 5 1 6 4 7 2 7 2 4 &   $5$ 
\end{tabular}
\hskip 10pt
\begin{tabular}{l|r|}
0 3 4 5 1 6 2 4 2 7 2 5 &   $5 $\\
0 3 4 5 1 6 4 6 2 7 2 4 &   $5 $\\
0 3 4 5 1 6 2 4 2 7 2 3 &   $5 $\\
0 3 4 5 2 6 4 7 5 7 1 2 &   $5 $\\
0 3 1 4 5 6 2 6 3 7 2 5 &   $5 $\\
0 3 4 5 2 5 6 7 1 2 4 6 &   $-5 $\\
0 3 1 4 5 6 2 7 3 5 2 6 &   $5 $\\
0 3 4 5 2 5 6 7 1 2 3 6 &   $-5 $\\
0 3 1 4 5 6 2 7 3 5 2 4 &   $5 $\\
0 3 4 5 2 6 6 7 3 7 1 2 &   $5 $\\
0 3 1 4 5 6 2 7 3 7 2 4 &   $5 $\\
0 3 4 5 5 6 6 7 1 2 2 3 &   $5 $\\
0 3 1 4 5 6 2 6 2 7 3 4 &   $5 $\\
0 3 4 5 1 2 2 6 4 7 2 4 &   $-5 $\\
0 3 1 4 2 5 3 6 2 7 2 3 &   $-5 $\\
0 3 4 5 2 6 6 7 1 2 2 6 &   $5 $\\
0 3 1 4 5 6 2 3 2 7 2 3 &   $-5 $\\
0 3 4 5 1 2 4 6 2 7 2 4 &   $-5 $\\
0 3 1 4 2 5 2 6 3 7 2 3 &   $-5$ 
\end{tabular}}

}

\smallskip
\begin{rem}
To establish 
the 
formula for 
the morphism $\Or$ that 
would be universal with respect to 
all cocycles $\gamma \in \ker \Id$, we are accumulating a sufficient number of pairs ($\Id$-\/cocycle $\gamma$, $\partial_\cP$-\/cocycle $\cQ$), in which $\cQ$ is built exactly 
from graphs that one obtains from orienting the graphs in~$\gamma$.
Let us remember that not only nontrivial cocycles (e.g., $\boldsymbol{\gamma}_3$,\ $\boldsymbol{\gamma}_5$,\ or $\boldsymbol{\gamma}_7$ from~\cite{JNMP17}, cf.~\cite{DolgushevRogersWillwacher,WillwacherGRT}) 
but also $\Id$-\/trivial, like $\delta_6$ on p.~\pageref{ExDifferential}, or even the `zero' non\/-\/oriented graphs are 
suited for this purpose:
e.g., a unique $\Or(w_4)(\cP)\equiv 0$ constrains~$\Or$.
In every such case, the respective $\partial_\cP$-\/cocycle is obtained\footnote{%
The actually found $\partial_\cP$-\/cocycle $\cQ$ might differ from the value $\Or(\gamma)$ by $\partial_\cP$-\/trivial or improper terms, 
i.e.\ 
$\cQ = \Or(\gamma) + \partial_\cP(\cX) + \nabla(\cP,\schouten{\cP,\cP}) $
for some vector field $\cX$ realized by Kontsevich graphs and for some ``Leibniz'' bi\/-\/vector graphs 
$\nabla$ 
vanishing identically at every Poisson structure~$\cP$.}
by
solving the factorization problem $\schouten{\cP,\cQ(\cP)} \doteq 0$ via $\schouten{\cP,\cP} = 0$.
The formula 
of the orientation morphism $\Or$ will be the object of another paper.
\end{rem} 

{\small
\noindent\textbf{Acknowledgements.}
The authors thank M.~Kontsevich and T.~Willwacher for recalling the existence of the orientation morphism~$\Or$.
A.V.K.\ thanks the organizers of international workshop SQS'17 (July~31 -- August~5, 2017 at JINR Dubna, Russia) for 
discussions.%
\footnote{As soon as the expression of $167$ Kontsevich 
graph coefficients in $\cQ_5$ via 
the $91$ 
integer parameters was obtained, 
the linear system 
in factorization $\schouten{\cP, \cQ_5(\cP)} = \Diamond(\cP, \schouten{\cP,\cP})$ for the
pentagon\/-\/wheel flow $\dot{\cP} = \cQ_5(\cP)$ was solved 
independently by A.~
Steel (Sydney) using the Markowitz pivoting run in \textsc{Magma}.
The flow components $\cQ_5$ of all the known solutions $(\cQ_5, \Diamond_5)$ match 
identically.
(For the flow $\dot{\cP} = \cQ_5(\cP) = \Or(\boldsymbol{\gamma}_5)(\cP)$, uniqueness is not claimed for the operator $\Diamond$ in the r.-h.s.\ of the factorization.)%
}

}


\end{minipage}%
]

\onecolumn
\setcounter{page}{1}
\renewcommand{\thepage}{\roman{page}}
\appendix
\section{The pentagon\/-\/wheel flow
: analytic formula}
\label{AppFormula}
\noindent%
Here is the value $\cQ_5(\cP)(f,g)$ of bi\/-\/vector~$\cQ_5$ at two functions~$f,g$:
\begin{align*}
10 \partial_{t} \partial_{m} \partial_{k} \cP^{ij} \partial_{p} \cP^{k\ell} \partial_{v} \partial_{r} \partial_{\ell} \cP^{mn} \partial_{n} \cP^{pq} \partial_{q} \cP^{rs} \partial_{s} \cP^{tv} \partial_{i} f \partial_{j} g \\
-10 \partial_{p} \partial_{m} \partial_{k} \cP^{ij} \partial_{t} \cP^{k\ell} \partial_{v} \partial_{r} \partial_{\ell} \cP^{mn} \partial_{n} \cP^{pq} \partial_{q} \cP^{rs} \partial_{s} \cP^{tv} \partial_{i} f \partial_{j} g \\
+10 \partial_{r} \partial_{m} \cP^{ij} \partial_{t} \partial_{j} \cP^{k\ell} \partial_{v} \partial_{s} \partial_{\ell} \cP^{mn} \partial_{n} \cP^{pq} \partial_{p} \cP^{rs} \partial_{q} \cP^{tv} \partial_{i} f \partial_{k} g \\
-10 \partial_{r} \partial_{n} \cP^{ij} \partial_{t} \partial_{s} \partial_{j} \cP^{k\ell} \partial_{v} \partial_{k} \cP^{mn} \partial_{\ell} \cP^{pq} \partial_{p} \cP^{rs} \partial_{q} \cP^{tv} \partial_{i} f \partial_{m} g \\
+10 \partial_{p} \partial_{m} \cP^{ij} \partial_{t} \partial_{j} \cP^{k\ell} \partial_{v} \partial_{r} \partial_{\ell} \cP^{mn} \partial_{n} \cP^{pq} \partial_{q} \cP^{rs} \partial_{s} \cP^{tv} \partial_{i} f \partial_{k} g \\
-10 \partial_{t} \partial_{n} \cP^{ij} \partial_{v} \partial_{r} \partial_{j} \cP^{k\ell} \partial_{p} \partial_{k} \cP^{mn} \partial_{\ell} \cP^{pq} \partial_{q} \cP^{rs} \partial_{s} \cP^{tv} \partial_{i} f \partial_{m} g \\
-10 \partial_{t} \partial_{m} \cP^{ij} \partial_{p} \partial_{j} \cP^{k\ell} \partial_{v} \partial_{r} \partial_{\ell} \cP^{mn} \partial_{n} \cP^{pq} \partial_{q} \cP^{rs} \partial_{s} \cP^{tv} \partial_{i} f \partial_{k} g \\
-10 \partial_{p} \partial_{n} \cP^{ij} \partial_{t} \partial_{r} \partial_{j} \cP^{k\ell} \partial_{v} \partial_{k} \cP^{mn} \partial_{\ell} \cP^{pq} \partial_{q} \cP^{rs} \partial_{s} \cP^{tv} \partial_{i} f \partial_{m} g \\
-10 \partial_{t} \partial_{p} \cP^{ij} \partial_{q} \partial_{j} \cP^{k\ell} \partial_{\ell} \cP^{mn} \partial_{v} \partial_{r} \partial_{m} \cP^{pq} \partial_{n} \cP^{rs} \partial_{s} \cP^{tv} \partial_{i} f \partial_{k} g \\
+10 \partial_{s} \partial_{m} \cP^{ij} \partial_{j} \cP^{k\ell} \partial_{t} \partial_{p} \partial_{k} \cP^{mn} \partial_{\ell} \cP^{pq} \partial_{v} \partial_{n} \cP^{rs} \partial_{q} \cP^{tv} \partial_{i} f \partial_{r} g \\
+10 \partial_{t} \partial_{p} \cP^{ij} \partial_{j} \cP^{k\ell} \partial_{q} \partial_{k} \cP^{mn} \partial_{v} \partial_{r} \partial_{\ell} \cP^{pq} \partial_{n} \cP^{rs} \partial_{s} \cP^{tv} \partial_{i} f \partial_{m} g \\
-10 \partial_{t} \partial_{m} \cP^{ij} \partial_{j} \cP^{k\ell} \partial_{v} \partial_{p} \partial_{k} \cP^{mn} \partial_{\ell} \cP^{pq} \partial_{q} \partial_{n} \cP^{rs} \partial_{s} \cP^{tv} \partial_{i} f \partial_{r} g \\
-10 \partial_{r} \partial_{m} \cP^{ij} \partial_{j} \cP^{k\ell} \partial_{v} \partial_{p} \partial_{k} \cP^{mn} \partial_{\ell} \cP^{pq} \partial_{n} \cP^{rs} \partial_{s} \partial_{q} \cP^{tv} \partial_{i} f \partial_{t} g \\
+10 \partial_{t} \partial_{p} \cP^{ij} \partial_{v} \partial_{r} \partial_{j} \cP^{k\ell} \partial_{q} \partial_{k} \cP^{mn} \partial_{\ell} \cP^{pq} \partial_{n} \cP^{rs} \partial_{s} \cP^{tv} \partial_{i} f \partial_{m} g \\
-10 \partial_{t} \partial_{m} \cP^{ij} \partial_{v} \partial_{s} \partial_{j} \cP^{k\ell} \partial_{k} \cP^{mn} \partial_{\ell} \cP^{pq} \partial_{p} \partial_{n} \cP^{rs} \partial_{q} \cP^{tv} \partial_{i} f \partial_{r} g \\
+10 \partial_{p} \partial_{m} \cP^{ij} \partial_{t} \partial_{s} \partial_{j} \cP^{k\ell} \partial_{k} \cP^{mn} \partial_{\ell} \cP^{pq} \partial_{v} \partial_{n} \cP^{rs} \partial_{q} \cP^{tv} \partial_{i} f \partial_{r} g \\
-10 \partial_{t} \partial_{r} \cP^{ij} \partial_{v} \partial_{s} \partial_{j} \cP^{k\ell} \partial_{p} \partial_{k} \cP^{mn} \partial_{\ell} \cP^{pq} \partial_{n} \cP^{rs} \partial_{q} \cP^{tv} \partial_{i} f \partial_{m} g \\
+10 \partial_{r} \partial_{p} \cP^{ij} \partial_{j} \cP^{k\ell} \partial_{t} \partial_{k} \cP^{mn} \partial_{v} \partial_{n} \partial_{\ell} \cP^{pq} \partial_{q} \cP^{rs} \partial_{s} \cP^{tv} \partial_{i} f \partial_{m} g \\
+10 \partial_{r} \partial_{p} \cP^{ij} \partial_{t} \partial_{s} \partial_{j} \cP^{k\ell} \partial_{v} \partial_{k} \cP^{mn} \partial_{\ell} \cP^{pq} \partial_{n} \cP^{rs} \partial_{q} \cP^{tv} \partial_{i} f \partial_{m} g \\
+10 \partial_{t} \partial_{p} \cP^{ij} \partial_{j} \cP^{k\ell} \partial_{r} \partial_{k} \cP^{mn} \partial_{v} \partial_{n} \partial_{\ell} \cP^{pq} \partial_{q} \cP^{rs} \partial_{s} \cP^{tv} \partial_{i} f \partial_{m} g \\
-10 \partial_{t} \partial_{n} \cP^{ij} \partial_{j} \cP^{k\ell} \partial_{v} \partial_{r} \partial_{p} \partial_{k} \cP^{mn} \partial_{\ell} \cP^{pq} \partial_{q} \cP^{rs} \partial_{s} \cP^{tv} \partial_{i} f \partial_{m} g \\
-10 \partial_{t} \partial_{r} \partial_{p} \partial_{m} \cP^{ij} \partial_{v} \partial_{j} \cP^{k\ell} \partial_{\ell} \cP^{mn} \partial_{n} \cP^{pq} \partial_{q} \cP^{rs} \partial_{s} \cP^{tv} \partial_{i} f \partial_{k} g \\
-10 \partial_{t} \partial_{m} \cP^{ij} \partial_{v} \partial_{r} \partial_{p} \partial_{j} \cP^{k\ell} \partial_{\ell} \cP^{mn} \partial_{n} \cP^{pq} \partial_{q} \cP^{rs} \partial_{s} \cP^{tv} \partial_{i} f \partial_{k} g \\
-10 \partial_{t} \partial_{r} \partial_{p} \partial_{n} \cP^{ij} \partial_{j} \cP^{k\ell} \partial_{v} \partial_{k} \cP^{mn} \partial_{\ell} \cP^{pq} \partial_{q} \cP^{rs} \partial_{s} \cP^{tv} \partial_{i} f \partial_{m} g \\
-10 \partial_{t} \partial_{p} \cP^{ij} \partial_{v} \partial_{r} \partial_{q} \partial_{j} \cP^{k\ell} \partial_{\ell} \cP^{mn} \partial_{m} \cP^{pq} \partial_{n} \cP^{rs} \partial_{s} \cP^{tv} \partial_{i} f \partial_{k} g \\
+10 \partial_{t} \partial_{s} \partial_{p} \partial_{m} \cP^{ij} \partial_{j} \cP^{k\ell} \partial_{k} \cP^{mn} \partial_{\ell} \cP^{pq} \partial_{v} \partial_{n} \cP^{rs} \partial_{q} \cP^{tv} \partial_{i} f \partial_{r} g \\
+2 \partial_{t} \partial_{r} \partial_{p} \partial_{m} \partial_{k} \cP^{ij} \partial_{v} \cP^{k\ell} \partial_{\ell} \cP^{mn} \partial_{n} \cP^{pq} \partial_{q} \cP^{rs} \partial_{s} \cP^{tv} \partial_{i} f \partial_{j} g \\
-5 \partial_{p} \partial_{m} \partial_{k} \cP^{ij} \partial_{t} \partial_{r} \cP^{k\ell} \partial_{v} \partial_{\ell} \cP^{mn} \partial_{n} \cP^{pq} \partial_{q} \cP^{rs} \partial_{s} \cP^{tv} \partial_{i} f \partial_{j} g \\
+5 \partial_{t} \partial_{m} \partial_{k} \cP^{ij} \partial_{r} \partial_{p} \cP^{k\ell} \partial_{v} \partial_{\ell} \cP^{mn} \partial_{n} \cP^{pq} \partial_{q} \cP^{rs} \partial_{s} \cP^{tv} \partial_{i} f \partial_{j} g \\
-5 \partial_{p} \partial_{m} \partial_{k} \cP^{ij} \partial_{r} \cP^{k\ell} \partial_{t} \partial_{\ell} \cP^{mn} \partial_{v} \partial_{n} \cP^{pq} \partial_{q} \cP^{rs} \partial_{s} \cP^{tv} \partial_{i} f \partial_{j} g \\
-5 \partial_{p} \partial_{m} \partial_{k} \cP^{ij} \partial_{t} \cP^{k\ell} \partial_{r} \partial_{\ell} \cP^{mn} \partial_{v} \partial_{n} \cP^{pq} \partial_{q} \cP^{rs} \partial_{s} \cP^{tv} \partial_{i} f \partial_{j} g \\
+5 \partial_{t} \partial_{p} \cP^{ij} \partial_{v} \partial_{j} \cP^{k\ell} \partial_{\ell} \cP^{mn} \partial_{r} \partial_{m} \cP^{pq} \partial_{n} \cP^{rs} \partial_{s} \partial_{q} \cP^{tv} \partial_{i} f \partial_{k} g \\
+5 \partial_{v} \partial_{r} \cP^{ij} \partial_{j} \cP^{k\ell} \partial_{k} \cP^{mn} \partial_{m} \partial_{\ell} \cP^{pq} \partial_{p} \partial_{n} \cP^{rs} \partial_{s} \partial_{q} \cP^{tv} \partial_{i} f \partial_{t} g 
\end{align*}
\begin{align*}
+5 \partial_{r} \partial_{m} \cP^{ij} \partial_{t} \partial_{j} \cP^{k\ell} \partial_{s} \partial_{\ell} \cP^{mn} \partial_{n} \cP^{pq} \partial_{v} \partial_{p} \cP^{rs} \partial_{q} \cP^{tv} \partial_{i} f \partial_{k} g \\
-5 \partial_{r} \partial_{n} \cP^{ij} \partial_{t} \partial_{j} \cP^{k\ell} \partial_{v} \partial_{k} \cP^{mn} \partial_{\ell} \cP^{pq} \partial_{p} \cP^{rs} \partial_{s} \partial_{q} \cP^{tv} \partial_{i} f \partial_{m} g \\
+5 \partial_{p} \partial_{m} \cP^{ij} \partial_{r} \partial_{j} \cP^{k\ell} \partial_{t} \partial_{\ell} \cP^{mn} \partial_{v} \partial_{n} \cP^{pq} \partial_{q} \cP^{rs} \partial_{s} \cP^{tv} \partial_{i} f \partial_{k} g \\
-5 \partial_{r} \partial_{n} \cP^{ij} \partial_{t} \partial_{j} \cP^{k\ell} \partial_{p} \partial_{k} \cP^{mn} \partial_{v} \partial_{\ell} \cP^{pq} \partial_{q} \cP^{rs} \partial_{s} \cP^{tv} \partial_{i} f \partial_{m} g \\
+5 \partial_{p} \partial_{m} \cP^{ij} \partial_{t} \partial_{j} \cP^{k\ell} \partial_{r} \partial_{\ell} \cP^{mn} \partial_{v} \partial_{n} \cP^{pq} \partial_{q} \cP^{rs} \partial_{s} \cP^{tv} \partial_{i} f \partial_{k} g \\
-5 \partial_{t} \partial_{n} \cP^{ij} \partial_{r} \partial_{j} \cP^{k\ell} \partial_{p} \partial_{k} \cP^{mn} \partial_{v} \partial_{\ell} \cP^{pq} \partial_{q} \cP^{rs} \partial_{s} \cP^{tv} \partial_{i} f \partial_{m} g \\
+5 \partial_{r} \partial_{n} \cP^{ij} \partial_{s} \partial_{j} \cP^{k\ell} \partial_{t} \partial_{k} \cP^{mn} \partial_{\ell} \cP^{pq} \partial_{v} \partial_{p} \cP^{rs} \partial_{q} \cP^{tv} \partial_{i} f \partial_{m} g \\
+5 \partial_{r} \partial_{m} \cP^{ij} \partial_{t} \partial_{j} \cP^{k\ell} \partial_{v} \partial_{\ell} \cP^{mn} \partial_{n} \cP^{pq} \partial_{p} \cP^{rs} \partial_{s} \partial_{q} \cP^{tv} \partial_{i} f \partial_{k} g \\
+5 \partial_{p} \partial_{n} \cP^{ij} \partial_{t} \partial_{j} \cP^{k\ell} \partial_{r} \partial_{k} \cP^{mn} \partial_{v} \partial_{\ell} \cP^{pq} \partial_{q} \cP^{rs} \partial_{s} \cP^{tv} \partial_{i} f \partial_{m} g \\
-5 \partial_{r} \partial_{m} \cP^{ij} \partial_{p} \partial_{j} \cP^{k\ell} \partial_{t} \partial_{\ell} \cP^{mn} \partial_{v} \partial_{n} \cP^{pq} \partial_{q} \cP^{rs} \partial_{s} \cP^{tv} \partial_{i} f \partial_{k} g \\
+5 \partial_{p} \partial_{n} \cP^{ij} \partial_{r} \partial_{j} \cP^{k\ell} \partial_{t} \partial_{k} \cP^{mn} \partial_{v} \partial_{\ell} \cP^{pq} \partial_{q} \cP^{rs} \partial_{s} \cP^{tv} \partial_{i} f \partial_{m} g \\
-5 \partial_{t} \partial_{m} \cP^{ij} \partial_{p} \partial_{j} \cP^{k\ell} \partial_{r} \partial_{\ell} \cP^{mn} \partial_{v} \partial_{n} \cP^{pq} \partial_{q} \cP^{rs} \partial_{s} \cP^{tv} \partial_{i} f \partial_{k} g \\
+5 \partial_{s} \partial_{m} \cP^{ij} \partial_{t} \partial_{j} \cP^{k\ell} \partial_{v} \partial_{k} \cP^{mn} \partial_{\ell} \cP^{pq} \partial_{p} \partial_{n} \cP^{rs} \partial_{q} \cP^{tv} \partial_{i} f \partial_{r} g \\
+5 \partial_{r} \partial_{p} \cP^{ij} \partial_{q} \partial_{j} \cP^{k\ell} \partial_{t} \partial_{\ell} \cP^{mn} \partial_{v} \partial_{m} \cP^{pq} \partial_{n} \cP^{rs} \partial_{s} \cP^{tv} \partial_{i} f \partial_{k} g \\
+5 \partial_{s} \partial_{m} \cP^{ij} \partial_{t} \partial_{j} \cP^{k\ell} \partial_{p} \partial_{k} \cP^{mn} \partial_{\ell} \cP^{pq} \partial_{v} \partial_{n} \cP^{rs} \partial_{q} \cP^{tv} \partial_{i} f \partial_{r} g \\
-5 \partial_{t} \partial_{p} \cP^{ij} \partial_{q} \partial_{j} \cP^{k\ell} \partial_{v} \partial_{\ell} \cP^{mn} \partial_{r} \partial_{m} \cP^{pq} \partial_{n} \cP^{rs} \partial_{s} \cP^{tv} \partial_{i} f \partial_{k} g \\
-5 \partial_{r} \partial_{n} \cP^{ij} \partial_{j} \cP^{k\ell} \partial_{t} \partial_{s} \partial_{k} \cP^{mn} \partial_{\ell} \cP^{pq} \partial_{v} \partial_{p} \cP^{rs} \partial_{q} \cP^{tv} \partial_{i} f \partial_{m} g \\
-5 \partial_{t} \partial_{r} \partial_{m} \cP^{ij} \partial_{s} \partial_{j} \cP^{k\ell} \partial_{\ell} \cP^{mn} \partial_{n} \cP^{pq} \partial_{v} \partial_{p} \cP^{rs} \partial_{q} \cP^{tv} \partial_{i} f \partial_{k} g \\
-5 \partial_{t} \partial_{p} \cP^{ij} \partial_{v} \partial_{q} \partial_{j} \cP^{k\ell} \partial_{\ell} \cP^{mn} \partial_{r} \partial_{m} \cP^{pq} \partial_{n} \cP^{rs} \partial_{s} \cP^{tv} \partial_{i} f \partial_{k} g \\
+5 \partial_{t} \partial_{s} \partial_{m} \cP^{ij} \partial_{j} \cP^{k\ell} \partial_{p} \partial_{k} \cP^{mn} \partial_{\ell} \cP^{pq} \partial_{v} \partial_{n} \cP^{rs} \partial_{q} \cP^{tv} \partial_{i} f \partial_{r} g \\
+5 \partial_{r} \partial_{m} \cP^{ij} \partial_{t} \partial_{s} \partial_{j} \cP^{k\ell} \partial_{v} \partial_{\ell} \cP^{mn} \partial_{n} \cP^{pq} \partial_{p} \cP^{rs} \partial_{q} \cP^{tv} \partial_{i} f \partial_{k} g \\
-5 \partial_{t} \partial_{r} \partial_{n} \cP^{ij} \partial_{s} \partial_{j} \cP^{k\ell} \partial_{v} \partial_{k} \cP^{mn} \partial_{\ell} \cP^{pq} \partial_{p} \cP^{rs} \partial_{q} \cP^{tv} \partial_{i} f \partial_{m} g \\
-5 \partial_{t} \partial_{m} \cP^{ij} \partial_{v} \partial_{p} \partial_{j} \cP^{k\ell} \partial_{r} \partial_{\ell} \cP^{mn} \partial_{n} \cP^{pq} \partial_{q} \cP^{rs} \partial_{s} \cP^{tv} \partial_{i} f \partial_{k} g \\
-5 \partial_{t} \partial_{p} \partial_{n} \cP^{ij} \partial_{r} \partial_{j} \cP^{k\ell} \partial_{v} \partial_{k} \cP^{mn} \partial_{\ell} \cP^{pq} \partial_{q} \cP^{rs} \partial_{s} \cP^{tv} \partial_{i} f \partial_{m} g \\
-5 \partial_{r} \partial_{m} \cP^{ij} \partial_{t} \partial_{s} \partial_{j} \cP^{k\ell} \partial_{\ell} \cP^{mn} \partial_{n} \cP^{pq} \partial_{v} \partial_{p} \cP^{rs} \partial_{q} \cP^{tv} \partial_{i} f \partial_{k} g \\
-5 \partial_{t} \partial_{r} \partial_{n} \cP^{ij} \partial_{j} \cP^{k\ell} \partial_{s} \partial_{k} \cP^{mn} \partial_{\ell} \cP^{pq} \partial_{v} \partial_{p} \cP^{rs} \partial_{q} \cP^{tv} \partial_{i} f \partial_{m} g \\
-5 \partial_{r} \partial_{n} \cP^{ij} \partial_{t} \partial_{j} \cP^{k\ell} \partial_{v} \partial_{s} \partial_{k} \cP^{mn} \partial_{\ell} \cP^{pq} \partial_{p} \cP^{rs} \partial_{q} \cP^{tv} \partial_{i} f \partial_{m} g \\
+5 \partial_{t} \partial_{r} \partial_{m} \cP^{ij} \partial_{s} \partial_{j} \cP^{k\ell} \partial_{v} \partial_{\ell} \cP^{mn} \partial_{n} \cP^{pq} \partial_{p} \cP^{rs} \partial_{q} \cP^{tv} \partial_{i} f \partial_{k} g \\
-5 \partial_{t} \partial_{n} \cP^{ij} \partial_{r} \partial_{j} \cP^{k\ell} \partial_{v} \partial_{p} \partial_{k} \cP^{mn} \partial_{\ell} \cP^{pq} \partial_{q} \cP^{rs} \partial_{s} \cP^{tv} \partial_{i} f \partial_{m} g \\
-5 \partial_{t} \partial_{p} \partial_{m} \cP^{ij} \partial_{v} \partial_{j} \cP^{k\ell} \partial_{r} \partial_{\ell} \cP^{mn} \partial_{n} \cP^{pq} \partial_{q} \cP^{rs} \partial_{s} \cP^{tv} \partial_{i} f \partial_{k} g \\
-5 \partial_{r} \partial_{m} \partial_{k} \cP^{ij} \partial_{t} \cP^{k\ell} \partial_{v} \partial_{\ell} \cP^{mn} \partial_{n} \cP^{pq} \partial_{p} \cP^{rs} \partial_{s} \partial_{q} \cP^{tv} \partial_{i} f \partial_{j} g \\
+5 \partial_{r} \partial_{m} \partial_{k} \cP^{ij} \partial_{p} \cP^{k\ell} \partial_{t} \partial_{\ell} \cP^{mn} \partial_{v} \partial_{n} \cP^{pq} \partial_{q} \cP^{rs} \partial_{s} \cP^{tv} \partial_{i} f \partial_{j} g \\
+5 \partial_{t} \partial_{m} \partial_{k} \cP^{ij} \partial_{p} \cP^{k\ell} \partial_{r} \partial_{\ell} \cP^{mn} \partial_{v} \partial_{n} \cP^{pq} \partial_{q} \cP^{rs} \partial_{s} \cP^{tv} \partial_{i} f \partial_{j} g \\
+5 \partial_{r} \partial_{m} \partial_{k} \cP^{ij} \partial_{t} \partial_{p} \cP^{k\ell} \partial_{\ell} \cP^{mn} \partial_{v} \partial_{n} \cP^{pq} \partial_{q} \cP^{rs} \partial_{s} \cP^{tv} \partial_{i} f \partial_{j} g \\
+5 \partial_{t} \partial_{m} \partial_{k} \cP^{ij} \partial_{r} \partial_{p} \cP^{k\ell} \partial_{\ell} \cP^{mn} \partial_{v} \partial_{n} \cP^{pq} \partial_{q} \cP^{rs} \partial_{s} \cP^{tv} \partial_{i} f \partial_{j} g \\
-5 \partial_{r} \partial_{n} \cP^{ij} \partial_{j} \cP^{k\ell} \partial_{t} \partial_{p} \partial_{k} \cP^{mn} \partial_{v} \partial_{\ell} \cP^{pq} \partial_{q} \cP^{rs} \partial_{s} \cP^{tv} \partial_{i} f \partial_{m} g 
\end{align*}
\begin{align*}
+5 \partial_{t} \partial_{p} \partial_{m} \cP^{ij} \partial_{r} \partial_{j} \cP^{k\ell} \partial_{\ell} \cP^{mn} \partial_{v} \partial_{n} \cP^{pq} \partial_{q} \cP^{rs} \partial_{s} \cP^{tv} \partial_{i} f \partial_{k} g \\
-5 \partial_{t} \partial_{n} \cP^{ij} \partial_{j} \cP^{k\ell} \partial_{r} \partial_{p} \partial_{k} \cP^{mn} \partial_{v} \partial_{\ell} \cP^{pq} \partial_{q} \cP^{rs} \partial_{s} \cP^{tv} \partial_{i} f \partial_{m} g \\
+5 \partial_{r} \partial_{p} \partial_{m} \cP^{ij} \partial_{t} \partial_{j} \cP^{k\ell} \partial_{\ell} \cP^{mn} \partial_{v} \partial_{n} \cP^{pq} \partial_{q} \cP^{rs} \partial_{s} \cP^{tv} \partial_{i} f \partial_{k} g \\
-5 \partial_{t} \partial_{p} \cP^{ij} \partial_{v} \partial_{r} \partial_{j} \cP^{k\ell} \partial_{\ell} \cP^{mn} \partial_{m} \cP^{pq} \partial_{q} \partial_{n} \cP^{rs} \partial_{s} \cP^{tv} \partial_{i} f \partial_{k} g \\
-5 \partial_{v} \partial_{r} \partial_{m} \cP^{ij} \partial_{j} \cP^{k\ell} \partial_{p} \partial_{k} \cP^{mn} \partial_{\ell} \cP^{pq} \partial_{n} \cP^{rs} \partial_{s} \partial_{q} \cP^{tv} \partial_{i} f \partial_{t} g \\
-5 \partial_{r} \partial_{p} \cP^{ij} \partial_{t} \partial_{q} \partial_{j} \cP^{k\ell} \partial_{v} \partial_{\ell} \cP^{mn} \partial_{m} \cP^{pq} \partial_{n} \cP^{rs} \partial_{s} \cP^{tv} \partial_{i} f \partial_{k} g \\
-5 \partial_{t} \partial_{s} \partial_{m} \cP^{ij} \partial_{v} \partial_{j} \cP^{k\ell} \partial_{k} \cP^{mn} \partial_{\ell} \cP^{pq} \partial_{p} \partial_{n} \cP^{rs} \partial_{q} \cP^{tv} \partial_{i} f \partial_{r} g \\
-5 \partial_{t} \partial_{p} \cP^{ij} \partial_{r} \partial_{q} \partial_{j} \cP^{k\ell} \partial_{v} \partial_{\ell} \cP^{mn} \partial_{m} \cP^{pq} \partial_{n} \cP^{rs} \partial_{s} \cP^{tv} \partial_{i} f \partial_{k} g \\
+5 \partial_{s} \partial_{p} \partial_{m} \cP^{ij} \partial_{t} \partial_{j} \cP^{k\ell} \partial_{k} \cP^{mn} \partial_{\ell} \cP^{pq} \partial_{v} \partial_{n} \cP^{rs} \partial_{q} \cP^{tv} \partial_{i} f \partial_{r} g \\
-5 \partial_{r} \partial_{m} \cP^{ij} \partial_{t} \partial_{p} \partial_{j} \cP^{k\ell} \partial_{\ell} \cP^{mn} \partial_{v} \partial_{n} \cP^{pq} \partial_{q} \cP^{rs} \partial_{s} \cP^{tv} \partial_{i} f \partial_{k} g \\
+5 \partial_{t} \partial_{p} \partial_{n} \cP^{ij} \partial_{j} \cP^{k\ell} \partial_{r} \partial_{k} \cP^{mn} \partial_{v} \partial_{\ell} \cP^{pq} \partial_{q} \cP^{rs} \partial_{s} \cP^{tv} \partial_{i} f \partial_{m} g \\
-5 \partial_{t} \partial_{m} \cP^{ij} \partial_{r} \partial_{p} \partial_{j} \cP^{k\ell} \partial_{\ell} \cP^{mn} \partial_{v} \partial_{n} \cP^{pq} \partial_{q} \cP^{rs} \partial_{s} \cP^{tv} \partial_{i} f \partial_{k} g \\
+5 \partial_{r} \partial_{p} \partial_{n} \cP^{ij} \partial_{j} \cP^{k\ell} \partial_{t} \partial_{k} \cP^{mn} \partial_{v} \partial_{\ell} \cP^{pq} \partial_{q} \cP^{rs} \partial_{s} \cP^{tv} \partial_{i} f \partial_{m} g \\
-5 \partial_{t} \partial_{s} \cP^{ij} \partial_{j} \cP^{k\ell} \partial_{k} \cP^{mn} \partial_{m} \partial_{\ell} \cP^{pq} \partial_{v} \partial_{p} \partial_{n} \cP^{rs} \partial_{q} \cP^{tv} \partial_{i} f \partial_{r} g \\
+5 \partial_{t} \partial_{r} \partial_{p} \cP^{ij} \partial_{v} \partial_{j} \cP^{k\ell} \partial_{\ell} \cP^{mn} \partial_{m} \cP^{pq} \partial_{q} \partial_{n} \cP^{rs} \partial_{s} \cP^{tv} \partial_{i} f \partial_{k} g \\
-5 \partial_{r} \partial_{p} \partial_{m} \partial_{k} \cP^{ij} \partial_{t} \cP^{k\ell} \partial_{v} \partial_{\ell} \cP^{mn} \partial_{n} \cP^{pq} \partial_{q} \cP^{rs} \partial_{s} \cP^{tv} \partial_{i} f \partial_{j} g \\
-5 \partial_{t} \partial_{p} \partial_{m} \partial_{k} \cP^{ij} \partial_{r} \cP^{k\ell} \partial_{\ell} \cP^{mn} \partial_{v} \partial_{n} \cP^{pq} \partial_{q} \cP^{rs} \partial_{s} \cP^{tv} \partial_{i} f \partial_{j} g \\
-5 \partial_{r} \partial_{p} \partial_{m} \partial_{k} \cP^{ij} \partial_{t} \cP^{k\ell} \partial_{\ell} \cP^{mn} \partial_{v} \partial_{n} \cP^{pq} \partial_{q} \cP^{rs} \partial_{s} \cP^{tv} \partial_{i} f \partial_{j} g \\
+5 \partial_{s} \partial_{m} \cP^{ij} \partial_{j} \cP^{k\ell} \partial_{t} \partial_{k} \cP^{mn} \partial_{\ell} \cP^{pq} \partial_{v} \partial_{p} \partial_{n} \cP^{rs} \partial_{q} \cP^{tv} \partial_{i} f \partial_{r} g \\
-5 \partial_{t} \partial_{r} \partial_{p} \cP^{ij} \partial_{q} \partial_{j} \cP^{k\ell} \partial_{\ell} \cP^{mn} \partial_{v} \partial_{m} \cP^{pq} \partial_{n} \cP^{rs} \partial_{s} \cP^{tv} \partial_{i} f \partial_{k} g \\
-5 \partial_{t} \partial_{n} \cP^{ij} \partial_{v} \partial_{j} \cP^{k\ell} \partial_{r} \partial_{p} \partial_{k} \cP^{mn} \partial_{\ell} \cP^{pq} \partial_{q} \cP^{rs} \partial_{s} \cP^{tv} \partial_{i} f \partial_{m} g \\
+5 \partial_{r} \partial_{p} \partial_{m} \cP^{ij} \partial_{t} \partial_{j} \cP^{k\ell} \partial_{v} \partial_{\ell} \cP^{mn} \partial_{n} \cP^{pq} \partial_{q} \cP^{rs} \partial_{s} \cP^{tv} \partial_{i} f \partial_{k} g \\
-5 \partial_{p} \partial_{n} \cP^{ij} \partial_{t} \partial_{j} \cP^{k\ell} \partial_{v} \partial_{r} \partial_{k} \cP^{mn} \partial_{\ell} \cP^{pq} \partial_{q} \cP^{rs} \partial_{s} \cP^{tv} \partial_{i} f \partial_{m} g \\
-5 \partial_{t} \partial_{r} \partial_{m} \cP^{ij} \partial_{p} \partial_{j} \cP^{k\ell} \partial_{v} \partial_{\ell} \cP^{mn} \partial_{n} \cP^{pq} \partial_{q} \cP^{rs} \partial_{s} \cP^{tv} \partial_{i} f \partial_{k} g \\
-5 \partial_{t} \partial_{p} \cP^{ij} \partial_{r} \partial_{q} \partial_{j} \cP^{k\ell} \partial_{\ell} \cP^{mn} \partial_{v} \partial_{m} \cP^{pq} \partial_{n} \cP^{rs} \partial_{s} \cP^{tv} \partial_{i} f \partial_{k} g \\
+5 \partial_{s} \partial_{p} \partial_{m} \cP^{ij} \partial_{j} \cP^{k\ell} \partial_{t} \partial_{k} \cP^{mn} \partial_{\ell} \cP^{pq} \partial_{v} \partial_{n} \cP^{rs} \partial_{q} \cP^{tv} \partial_{i} f \partial_{r} g \\
-5 \partial_{t} \partial_{p} \cP^{ij} \partial_{v} \partial_{r} \partial_{j} \cP^{k\ell} \partial_{\ell} \cP^{mn} \partial_{m} \cP^{pq} \partial_{n} \cP^{rs} \partial_{s} \partial_{q} \cP^{tv} \partial_{i} f \partial_{k} g \\
-5 \partial_{v} \partial_{r} \partial_{m} \cP^{ij} \partial_{j} \cP^{k\ell} \partial_{k} \cP^{mn} \partial_{\ell} \cP^{pq} \partial_{p} \partial_{n} \cP^{rs} \partial_{s} \partial_{q} \cP^{tv} \partial_{i} f \partial_{t} g \\
-5 \partial_{t} \partial_{m} \cP^{ij} \partial_{r} \partial_{p} \partial_{j} \cP^{k\ell} \partial_{v} \partial_{\ell} \cP^{mn} \partial_{n} \cP^{pq} \partial_{q} \cP^{rs} \partial_{s} \cP^{tv} \partial_{i} f \partial_{k} g \\
-5 \partial_{r} \partial_{p} \partial_{n} \cP^{ij} \partial_{t} \partial_{j} \cP^{k\ell} \partial_{v} \partial_{k} \cP^{mn} \partial_{\ell} \cP^{pq} \partial_{q} \cP^{rs} \partial_{s} \cP^{tv} \partial_{i} f \partial_{m} g \\
+5 \partial_{p} \partial_{m} \cP^{ij} \partial_{t} \partial_{r} \partial_{j} \cP^{k\ell} \partial_{v} \partial_{\ell} \cP^{mn} \partial_{n} \cP^{pq} \partial_{q} \cP^{rs} \partial_{s} \cP^{tv} \partial_{i} f \partial_{k} g \\
-5 \partial_{t} \partial_{r} \partial_{n} \cP^{ij} \partial_{v} \partial_{j} \cP^{k\ell} \partial_{p} \partial_{k} \cP^{mn} \partial_{\ell} \cP^{pq} \partial_{q} \cP^{rs} \partial_{s} \cP^{tv} \partial_{i} f \partial_{m} g \\
-5 \partial_{t} \partial_{p} \cP^{ij} \partial_{j} \cP^{k\ell} \partial_{v} \partial_{k} \cP^{mn} \partial_{r} \partial_{\ell} \cP^{pq} \partial_{n} \cP^{rs} \partial_{s} \partial_{q} \cP^{tv} \partial_{i} f \partial_{m} g \\
-5 \partial_{t} \partial_{m} \cP^{ij} \partial_{j} \cP^{k\ell} \partial_{p} \partial_{k} \cP^{mn} \partial_{v} \partial_{\ell} \cP^{pq} \partial_{q} \partial_{n} \cP^{rs} \partial_{s} \cP^{tv} \partial_{i} f \partial_{r} g \\
+5 \partial_{r} \partial_{p} \cP^{ij} \partial_{j} \cP^{k\ell} \partial_{t} \partial_{k} \cP^{mn} \partial_{s} \partial_{\ell} \cP^{pq} \partial_{v} \partial_{n} \cP^{rs} \partial_{q} \cP^{tv} \partial_{i} f \partial_{m} g \\
-5 \partial_{t} \partial_{r} \cP^{ij} \partial_{v} \partial_{j} \cP^{k\ell} \partial_{p} \partial_{k} \cP^{mn} \partial_{s} \partial_{\ell} \cP^{pq} \partial_{n} \cP^{rs} \partial_{q} \cP^{tv} \partial_{i} f \partial_{m} g 
\end{align*}
\begin{align*}
+5 \partial_{r} \partial_{p} \cP^{ij} \partial_{j} \cP^{k\ell} \partial_{t} \partial_{k} \cP^{mn} \partial_{v} \partial_{\ell} \cP^{pq} \partial_{q} \partial_{n} \cP^{rs} \partial_{s} \cP^{tv} \partial_{i} f \partial_{m} g \\
+5 \partial_{r} \partial_{p} \cP^{ij} \partial_{t} \partial_{j} \cP^{k\ell} \partial_{v} \partial_{k} \cP^{mn} \partial_{\ell} \cP^{pq} \partial_{n} \cP^{rs} \partial_{s} \partial_{q} \cP^{tv} \partial_{i} f \partial_{m} g \\
-5 \partial_{t} \partial_{m} \cP^{ij} \partial_{v} \partial_{j} \cP^{k\ell} \partial_{k} \cP^{mn} \partial_{\ell} \cP^{pq} \partial_{p} \partial_{n} \cP^{rs} \partial_{s} \partial_{q} \cP^{tv} \partial_{i} f \partial_{r} g \\
+5 \partial_{t} \partial_{r} \cP^{ij} \partial_{s} \partial_{j} \cP^{k\ell} \partial_{v} \partial_{k} \cP^{mn} \partial_{n} \partial_{\ell} \cP^{pq} \partial_{p} \cP^{rs} \partial_{q} \cP^{tv} \partial_{i} f \partial_{m} g \\
-5 \partial_{t} \partial_{m} \cP^{ij} \partial_{v} \partial_{j} \cP^{k\ell} \partial_{p} \partial_{k} \cP^{mn} \partial_{\ell} \cP^{pq} \partial_{q} \partial_{n} \cP^{rs} \partial_{s} \cP^{tv} \partial_{i} f \partial_{r} g \\
+5 \partial_{t} \partial_{r} \cP^{ij} \partial_{j} \cP^{k\ell} \partial_{s} \partial_{k} \cP^{mn} \partial_{n} \partial_{\ell} \cP^{pq} \partial_{v} \partial_{p} \cP^{rs} \partial_{q} \cP^{tv} \partial_{i} f \partial_{m} g \\
+5 \partial_{p} \partial_{m} \cP^{ij} \partial_{t} \partial_{j} \cP^{k\ell} \partial_{k} \cP^{mn} \partial_{s} \partial_{\ell} \cP^{pq} \partial_{v} \partial_{n} \cP^{rs} \partial_{q} \cP^{tv} \partial_{i} f \partial_{r} g \\
+5 \partial_{t} \partial_{p} \cP^{ij} \partial_{r} \partial_{j} \cP^{k\ell} \partial_{v} \partial_{k} \cP^{mn} \partial_{\ell} \cP^{pq} \partial_{q} \partial_{n} \cP^{rs} \partial_{s} \cP^{tv} \partial_{i} f \partial_{m} g \\
-5 \partial_{p} \partial_{m} \cP^{ij} \partial_{t} \partial_{j} \cP^{k\ell} \partial_{v} \partial_{k} \cP^{mn} \partial_{\ell} \cP^{pq} \partial_{q} \partial_{n} \cP^{rs} \partial_{s} \cP^{tv} \partial_{i} f \partial_{r} g \\
-5 \partial_{t} \partial_{m} \cP^{ij} \partial_{j} \cP^{k\ell} \partial_{p} \partial_{k} \cP^{mn} \partial_{s} \partial_{\ell} \cP^{pq} \partial_{v} \partial_{n} \cP^{rs} \partial_{q} \cP^{tv} \partial_{i} f \partial_{r} g \\
+5 \partial_{r} \partial_{p} \cP^{ij} \partial_{t} \partial_{j} \cP^{k\ell} \partial_{v} \partial_{k} \cP^{mn} \partial_{s} \partial_{\ell} \cP^{pq} \partial_{n} \cP^{rs} \partial_{q} \cP^{tv} \partial_{i} f \partial_{m} g \\
+5 \partial_{t} \partial_{r} \cP^{ij} \partial_{j} \cP^{k\ell} \partial_{p} \partial_{k} \cP^{mn} \partial_{s} \partial_{\ell} \cP^{pq} \partial_{v} \partial_{n} \cP^{rs} \partial_{q} \cP^{tv} \partial_{i} f \partial_{m} g \\
-5 \partial_{t} \partial_{r} \cP^{ij} \partial_{v} \partial_{j} \cP^{k\ell} \partial_{p} \partial_{k} \cP^{mn} \partial_{\ell} \cP^{pq} \partial_{n} \cP^{rs} \partial_{s} \partial_{q} \cP^{tv} \partial_{i} f \partial_{m} g \\
+5 \partial_{t} \partial_{r} \cP^{ij} \partial_{j} \cP^{k\ell} \partial_{p} \partial_{k} \cP^{mn} \partial_{v} \partial_{\ell} \cP^{pq} \partial_{q} \partial_{n} \cP^{rs} \partial_{s} \cP^{tv} \partial_{i} f \partial_{m} g \\
-5 \partial_{t} \partial_{p} \cP^{ij} \partial_{r} \partial_{j} \cP^{k\ell} \partial_{v} \partial_{k} \cP^{mn} \partial_{s} \partial_{\ell} \cP^{pq} \partial_{n} \cP^{rs} \partial_{q} \cP^{tv} \partial_{i} f \partial_{m} g \\
-5 \partial_{t} \partial_{m} \cP^{ij} \partial_{j} \cP^{k\ell} \partial_{s} \partial_{k} \cP^{mn} \partial_{n} \partial_{\ell} \cP^{pq} \partial_{v} \partial_{p} \cP^{rs} \partial_{q} \cP^{tv} \partial_{i} f \partial_{r} g \\
-5 \partial_{r} \partial_{m} \cP^{ij} \partial_{v} \partial_{j} \cP^{k\ell} \partial_{p} \partial_{k} \cP^{mn} \partial_{\ell} \cP^{pq} \partial_{n} \cP^{rs} \partial_{s} \partial_{q} \cP^{tv} \partial_{i} f \partial_{t} g \\
-5 \partial_{t} \partial_{m} \cP^{ij} \partial_{s} \partial_{j} \cP^{k\ell} \partial_{k} \cP^{mn} \partial_{n} \partial_{\ell} \cP^{pq} \partial_{v} \partial_{p} \cP^{rs} \partial_{q} \cP^{tv} \partial_{i} f \partial_{r} g \\
+5 \partial_{t} \partial_{m} \cP^{ij} \partial_{j} \cP^{k\ell} \partial_{v} \partial_{k} \cP^{mn} \partial_{\ell} \cP^{pq} \partial_{p} \partial_{n} \cP^{rs} \partial_{s} \partial_{q} \cP^{tv} \partial_{i} f \partial_{r} g \\
+5 \partial_{t} \partial_{p} \cP^{ij} \partial_{v} \partial_{j} \cP^{k\ell} \partial_{q} \partial_{k} \cP^{mn} \partial_{r} \partial_{\ell} \cP^{pq} \partial_{n} \cP^{rs} \partial_{s} \cP^{tv} \partial_{i} f \partial_{m} g \\
-5 \partial_{r} \partial_{m} \cP^{ij} \partial_{j} \cP^{k\ell} \partial_{v} \partial_{k} \cP^{mn} \partial_{\ell} \cP^{pq} \partial_{p} \partial_{n} \cP^{rs} \partial_{s} \partial_{q} \cP^{tv} \partial_{i} f \partial_{t} g \\
+5 \partial_{t} \partial_{p} \cP^{ij} \partial_{r} \partial_{j} \cP^{k\ell} \partial_{q} \partial_{k} \cP^{mn} \partial_{v} \partial_{\ell} \cP^{pq} \partial_{n} \cP^{rs} \partial_{s} \cP^{tv} \partial_{i} f \partial_{m} g \\
+5 \partial_{p} \partial_{m} \cP^{ij} \partial_{s} \partial_{j} \cP^{k\ell} \partial_{t} \partial_{k} \cP^{mn} \partial_{\ell} \cP^{pq} \partial_{v} \partial_{n} \cP^{rs} \partial_{q} \cP^{tv} \partial_{i} f \partial_{r} g \\
-5 \partial_{t} \partial_{m} \cP^{ij} \partial_{s} \partial_{j} \cP^{k\ell} \partial_{v} \partial_{k} \cP^{mn} \partial_{\ell} \cP^{pq} \partial_{p} \partial_{n} \cP^{rs} \partial_{q} \cP^{tv} \partial_{i} f \partial_{r} g \\
-5 \partial_{r} \partial_{p} \cP^{ij} \partial_{s} \partial_{j} \cP^{k\ell} \partial_{t} \partial_{k} \cP^{mn} \partial_{\ell} \cP^{pq} \partial_{v} \partial_{n} \cP^{rs} \partial_{q} \cP^{tv} \partial_{i} f \partial_{m} g \\
+5 \partial_{t} \partial_{p} \cP^{ij} \partial_{v} \partial_{j} \cP^{k\ell} \partial_{r} \partial_{k} \cP^{mn} \partial_{n} \partial_{\ell} \cP^{pq} \partial_{q} \cP^{rs} \partial_{s} \cP^{tv} \partial_{i} f \partial_{m} g \\
-5 \partial_{r} \partial_{p} \cP^{ij} \partial_{t} \partial_{j} \cP^{k\ell} \partial_{v} \partial_{k} \cP^{mn} \partial_{n} \partial_{\ell} \cP^{pq} \partial_{q} \cP^{rs} \partial_{s} \cP^{tv} \partial_{i} f \partial_{m} g \\
-5 \partial_{t} \partial_{r} \cP^{ij} \partial_{s} \partial_{j} \cP^{k\ell} \partial_{p} \partial_{k} \cP^{mn} \partial_{\ell} \cP^{pq} \partial_{v} \partial_{n} \cP^{rs} \partial_{q} \cP^{tv} \partial_{i} f \partial_{m} g \\
-5 \partial_{r} \partial_{p} \cP^{ij} \partial_{j} \cP^{k\ell} \partial_{t} \partial_{q} \partial_{k} \cP^{mn} \partial_{v} \partial_{\ell} \cP^{pq} \partial_{n} \cP^{rs} \partial_{s} \cP^{tv} \partial_{i} f \partial_{m} g \\
-5 \partial_{t} \partial_{r} \partial_{p} \cP^{ij} \partial_{j} \cP^{k\ell} \partial_{v} \partial_{k} \cP^{mn} \partial_{\ell} \cP^{pq} \partial_{n} \cP^{rs} \partial_{s} \partial_{q} \cP^{tv} \partial_{i} f \partial_{m} g \\
+5 \partial_{t} \partial_{p} \cP^{ij} \partial_{j} \cP^{k\ell} \partial_{v} \partial_{q} \partial_{k} \cP^{mn} \partial_{r} \partial_{\ell} \cP^{pq} \partial_{n} \cP^{rs} \partial_{s} \cP^{tv} \partial_{i} f \partial_{m} g \\
-5 \partial_{t} \partial_{p} \partial_{m} \cP^{ij} \partial_{j} \cP^{k\ell} \partial_{v} \partial_{k} \cP^{mn} \partial_{\ell} \cP^{pq} \partial_{q} \partial_{n} \cP^{rs} \partial_{s} \cP^{tv} \partial_{i} f \partial_{r} g \\
+5 \partial_{t} \partial_{p} \cP^{ij} \partial_{j} \cP^{k\ell} \partial_{v} \partial_{r} \partial_{k} \cP^{mn} \partial_{n} \partial_{\ell} \cP^{pq} \partial_{q} \cP^{rs} \partial_{s} \cP^{tv} \partial_{i} f \partial_{m} g \\
+5 \partial_{t} \partial_{r} \partial_{p} \cP^{ij} \partial_{s} \partial_{j} \cP^{k\ell} \partial_{v} \partial_{k} \cP^{mn} \partial_{\ell} \cP^{pq} \partial_{n} \cP^{rs} \partial_{q} \cP^{tv} \partial_{i} f \partial_{m} g \\
+5 \partial_{r} \partial_{p} \cP^{ij} \partial_{t} \partial_{j} \cP^{k\ell} \partial_{v} \partial_{q} \partial_{k} \cP^{mn} \partial_{\ell} \cP^{pq} \partial_{n} \cP^{rs} \partial_{s} \cP^{tv} \partial_{i} f \partial_{m} g 
\end{align*}
\begin{align*}
-5 \partial_{t} \partial_{r} \partial_{p} \cP^{ij} \partial_{j} \cP^{k\ell} \partial_{v} \partial_{k} \cP^{mn} \partial_{\ell} \cP^{pq} \partial_{q} \partial_{n} \cP^{rs} \partial_{s} \cP^{tv} \partial_{i} f \partial_{m} g \\
+5 \partial_{t} \partial_{p} \cP^{ij} \partial_{r} \partial_{j} \cP^{k\ell} \partial_{v} \partial_{q} \partial_{k} \cP^{mn} \partial_{\ell} \cP^{pq} \partial_{n} \cP^{rs} \partial_{s} \cP^{tv} \partial_{i} f \partial_{m} g \\
+5 \partial_{t} \partial_{p} \partial_{m} \cP^{ij} \partial_{j} \cP^{k\ell} \partial_{k} \cP^{mn} \partial_{s} \partial_{\ell} \cP^{pq} \partial_{v} \partial_{n} \cP^{rs} \partial_{q} \cP^{tv} \partial_{i} f \partial_{r} g \\
-5 \partial_{t} \partial_{m} \cP^{ij} \partial_{s} \partial_{j} \cP^{k\ell} \partial_{k} \cP^{mn} \partial_{\ell} \cP^{pq} \partial_{v} \partial_{p} \partial_{n} \cP^{rs} \partial_{q} \cP^{tv} \partial_{i} f \partial_{r} g \\
+5 \partial_{t} \partial_{p} \partial_{m} \cP^{ij} \partial_{s} \partial_{j} \cP^{k\ell} \partial_{k} \cP^{mn} \partial_{\ell} \cP^{pq} \partial_{v} \partial_{n} \cP^{rs} \partial_{q} \cP^{tv} \partial_{i} f \partial_{r} g \\
-5 \partial_{t} \partial_{r} \cP^{ij} \partial_{s} \partial_{j} \cP^{k\ell} \partial_{v} \partial_{p} \partial_{k} \cP^{mn} \partial_{\ell} \cP^{pq} \partial_{n} \cP^{rs} \partial_{q} \cP^{tv} \partial_{i} f \partial_{m} g \\
-5 \partial_{t} \partial_{r} \partial_{p} \cP^{ij} \partial_{j} \cP^{k\ell} \partial_{v} \partial_{k} \cP^{mn} \partial_{n} \partial_{\ell} \cP^{pq} \partial_{q} \cP^{rs} \partial_{s} \cP^{tv} \partial_{i} f \partial_{m} g \\
+5 \partial_{t} \partial_{r} \cP^{ij} \partial_{j} \cP^{k\ell} \partial_{v} \partial_{p} \partial_{k} \cP^{mn} \partial_{\ell} \cP^{pq} \partial_{n} \cP^{rs} \partial_{s} \partial_{q} \cP^{tv} \partial_{i} f \partial_{m} g \\
+5 \partial_{t} \partial_{r} \partial_{p} \cP^{ij} \partial_{j} \cP^{k\ell} \partial_{q} \partial_{k} \cP^{mn} \partial_{v} \partial_{\ell} \cP^{pq} \partial_{n} \cP^{rs} \partial_{s} \cP^{tv} \partial_{i} f \partial_{m} g \\
+5 \partial_{t} \partial_{r} \cP^{ij} \partial_{j} \cP^{k\ell} \partial_{v} \partial_{p} \partial_{k} \cP^{mn} \partial_{\ell} \cP^{pq} \partial_{q} \partial_{n} \cP^{rs} \partial_{s} \cP^{tv} \partial_{i} f \partial_{m} g \\
+5 \partial_{t} \partial_{r} \partial_{p} \cP^{ij} \partial_{v} \partial_{j} \cP^{k\ell} \partial_{q} \partial_{k} \cP^{mn} \partial_{\ell} \cP^{pq} \partial_{n} \cP^{rs} \partial_{s} \cP^{tv} \partial_{i} f \partial_{m} g \\
+5 \partial_{v} \partial_{m} \cP^{ij} \partial_{j} \cP^{k\ell} \partial_{p} \partial_{k} \cP^{mn} \partial_{r} \partial_{\ell} \cP^{pq} \partial_{n} \cP^{rs} \partial_{s} \partial_{q} \cP^{tv} \partial_{i} f \partial_{t} g \\
+5 \partial_{t} \partial_{p} \cP^{ij} \partial_{r} \partial_{j} \cP^{k\ell} \partial_{\ell} \cP^{mn} \partial_{v} \partial_{m} \cP^{pq} \partial_{q} \partial_{n} \cP^{rs} \partial_{s} \cP^{tv} \partial_{i} f \partial_{k} g \\
-5 \partial_{s} \partial_{m} \cP^{ij} \partial_{j} \cP^{k\ell} \partial_{t} \partial_{k} \cP^{mn} \partial_{n} \partial_{\ell} \cP^{pq} \partial_{v} \partial_{p} \cP^{rs} \partial_{q} \cP^{tv} \partial_{i} f \partial_{r} g \\
+5 \partial_{t} \partial_{p} \cP^{ij} \partial_{r} \partial_{j} \cP^{k\ell} \partial_{\ell} \cP^{mn} \partial_{s} \partial_{m} \cP^{pq} \partial_{v} \partial_{n} \cP^{rs} \partial_{q} \cP^{tv} \partial_{i} f \partial_{k} g \\
-5 \partial_{s} \partial_{m} \cP^{ij} \partial_{t} \partial_{j} \cP^{k\ell} \partial_{k} \cP^{mn} \partial_{n} \partial_{\ell} \cP^{pq} \partial_{v} \partial_{p} \cP^{rs} \partial_{q} \cP^{tv} \partial_{i} f \partial_{r} g \\
+5 \partial_{t} \partial_{p} \cP^{ij} \partial_{r} \partial_{j} \cP^{k\ell} \partial_{v} \partial_{\ell} \cP^{mn} \partial_{s} \partial_{m} \cP^{pq} \partial_{n} \cP^{rs} \partial_{q} \cP^{tv} \partial_{i} f \partial_{k} g \\
+5 \partial_{v} \partial_{m} \cP^{ij} \partial_{r} \partial_{j} \cP^{k\ell} \partial_{k} \cP^{mn} \partial_{\ell} \cP^{pq} \partial_{p} \partial_{n} \cP^{rs} \partial_{s} \partial_{q} \cP^{tv} \partial_{i} f \partial_{t} g \\
+5 \partial_{t} \partial_{p} \cP^{ij} \partial_{r} \partial_{j} \cP^{k\ell} \partial_{v} \partial_{\ell} \cP^{mn} \partial_{m} \cP^{pq} \partial_{n} \cP^{rs} \partial_{s} \partial_{q} \cP^{tv} \partial_{i} f \partial_{k} g \\
+5 \partial_{t} \partial_{s} \cP^{ij} \partial_{v} \partial_{j} \cP^{k\ell} \partial_{k} \cP^{mn} \partial_{m} \partial_{\ell} \cP^{pq} \partial_{p} \partial_{n} \cP^{rs} \partial_{q} \cP^{tv} \partial_{i} f \partial_{r} g \\
+5 \partial_{r} \partial_{p} \cP^{ij} \partial_{t} \partial_{j} \cP^{k\ell} \partial_{v} \partial_{\ell} \cP^{mn} \partial_{m} \cP^{pq} \partial_{q} \partial_{n} \cP^{rs} \partial_{s} \cP^{tv} \partial_{i} f \partial_{k} g \\
-5 \partial_{t} \partial_{p} \partial_{n} \cP^{ij} \partial_{j} \cP^{k\ell} \partial_{v} \partial_{r} \partial_{k} \cP^{mn} \partial_{\ell} \cP^{pq} \partial_{q} \cP^{rs} \partial_{s} \cP^{tv} \partial_{i} f \partial_{m} g \\
-5 \partial_{t} \partial_{r} \partial_{m} \cP^{ij} \partial_{v} \partial_{p} \partial_{j} \cP^{k\ell} \partial_{\ell} \cP^{mn} \partial_{n} \cP^{pq} \partial_{q} \cP^{rs} \partial_{s} \cP^{tv} \partial_{i} f \partial_{k} g \\
+5 \partial_{t} \partial_{s} \partial_{m} \cP^{ij} \partial_{j} \cP^{k\ell} \partial_{k} \cP^{mn} \partial_{\ell} \cP^{pq} \partial_{v} \partial_{p} \partial_{n} \cP^{rs} \partial_{q} \cP^{tv} \partial_{i} f \partial_{r} g \\
-5 \partial_{t} \partial_{r} \partial_{p} \cP^{ij} \partial_{v} \partial_{q} \partial_{j} \cP^{k\ell} \partial_{\ell} \cP^{mn} \partial_{m} \cP^{pq} \partial_{n} \cP^{rs} \partial_{s} \cP^{tv} \partial_{i} f \partial_{k} g \\
-5 \partial_{t} \partial_{r} \partial_{n} \cP^{ij} \partial_{j} \cP^{k\ell} \partial_{v} \partial_{p} \partial_{k} \cP^{mn} \partial_{\ell} \cP^{pq} \partial_{q} \cP^{rs} \partial_{s} \cP^{tv} \partial_{i} f \partial_{m} g \\
-5 \partial_{t} \partial_{p} \partial_{m} \cP^{ij} \partial_{v} \partial_{r} \partial_{j} \cP^{k\ell} \partial_{\ell} \cP^{mn} \partial_{n} \cP^{pq} \partial_{q} \cP^{rs} \partial_{s} \cP^{tv} \partial_{i} f \partial_{k} g.
\end{align*}
In every term, the Einstein summation convention works for each repeated index (i.e.\ once upper and another time lower), the indices running from~$1$ to the dimension~$r<\infty$ of the affine Poisson manifold~$M^r$ at hand.

\end{document}